\documentclass[12pt]{article}
\usepackage{graphicx}
\usepackage{epsfig}
\usepackage{pst-plot,url,epsf}
\usepackage{t1enc}

\newcommand{\beq}{\begin{equation}}
\newcommand{\eeq}{\end{equation}}
\newcommand{\bea}{\begin{eqnarray}}
\newcommand{\eea}{\end{eqnarray}}

\newcommand{\epm}{e^+e^-}
\newcommand{\nn}{\nonumber}

\newcommand{\ra}{\rightarrow}
\def\earr{\end{array}}
\def\barr#1{\begin{array}{#1}}

\begin{document}
\thispagestyle{empty}
\begin{flushright}
Revised version\\
August 2013
\vspace*{1.5cm}
\end{flushright}
\begin{center}
{\LARGE\bf Anomalous $Wtb$ coupling at the LHC
}\\
\vspace*{2cm}
K. Ko\l odziej\footnote{E-mail: karol.kolodziej@us.edu.pl}\\[1cm]
{\small\it
Institute of Physics, University of Silesia\\ 
ul. Uniwersytecka 4, PL-40007 Katowice, Poland}\\
\vspace*{3.5cm}
{\bf Abstract}\\
\end{center}
Some distributions of the $\mu^-$ in the top quark
pair production reaction $pp \;\ra\; b u\bar{d}\;\bar{b} \mu^-\bar{\nu}_{\mu}$
at the LHC are calculated to leading order in the presence 
of the anomalous $Wtb$ coupling with operators of dimension up to 
five. The distributions in the transverse momentum, rapidity and 
cosine of the $\mu^-$ angle with respect to the beam in the laboratory frame 
and with respect to the reversed momentum of the $b$-quark in the
rest frame of $W$-boson are changed rather moderately by the anomalous 
$Wtb$ coupling. The distributions computed with the full set of leading order 
Feynman diagrams practically do not differ from those computed with the 
$t\bar t$ production diagrams, with typical acceptance cuts. This demonstrates
very little effect of the off resonance background contributions.

\vfill
\newpage

\section{Introduction.}

As the top quark is very heavy, it 
decays even before it can hadronize, predominantly into a 
$W$-boson and a quark, with the branching fraction of $t\to Wb$ close to 1,
which means that the decay at the leading order is
almost exclusively governed by the $Wtb$ coupling.  
Due to the almost immediate decay, the information about the top quark 
spin and couplings is passed to
its decay products without being obscured by the hadronization process
and it can be best gained from the analysis of differential cross sections,
in particular from the angular distribution of the lepton from the $W$-boson 
decay \cite{Jezabek}.
Therefore the top quark production processes are ideal for tests of
extensions of the standard model (SM)
that lead to modifications of the pure left-handed $Wtb$ coupling of SM.
This issue has been already extensively addressed in literature, see, e.g.,
\cite{kane}, \cite{Wtb}, \cite{GHR}, \cite{CK1}, \cite{tt6fwtb}. 
In \cite{GHR}, a decoupling theorem was proven,
which states that the angular distribution of the secondary lepton resulting
from a decay of the top quark produced in $\epm\to t\bar t$ receives no 
contribution from the anomalous $Wtb$ coupling
in the narrow width approximation (NWA). The theorem turned out to 
remain correct also in a more complicated case of the off 
shell top quark pair production in $\epm$
annihilation and decay in 6 fermion final states including
the non-double resonance background contributions, which was checked
by direct computation in \cite{tt6fwtb}.
It was also shown \cite{afbKK} that the anomalous 
$Wtb$ coupling cannot explain discrepancies between the forward-backward 
asymmetry (FBA) in the top quark pair production in high energy 
proton-antiproton collisions observed by the CDF \cite{afbCDF} 
and D0 \cite{afbD0} experiments at Tevatron and the SM expectations 
\cite{afbSM}.
The FBA in \cite{afbKK} was computed taking into account leading order 
cross sections of all the sub-processes of
quark-antiquark annihilation, which dominate the top 
quark pair production at the Tevatron, of the form
$q\bar{q} \;\ra\; b q\bar{q}'\;\bar{b} l\bar{\nu}_l$,
with $u\bar u$ and $d\bar d$ in the initial state and a single charged lepton 
in the final state. 
In spite of the fact that off resonance contributions to such reactions
are changed in the presence 
of the anomalous $Wtb$ coupling, as it substantially alters the
total top quark decay width, the rapidity distributions 
of the final state lepton remain almost unchanged. This can be considered
as another example of the decoupling theorem 
of \cite{GHR} that remains
valid in the off shell top quark pair production and decay
in quark-antiquark annihilation.

In the proton-proton collisions at the LHC, the top quarks are produced 
dominantly in pairs through
the underlying gluon-gluon fusion or the quark-antiquark annihilation processes
\bea
\label{ggtt}
gg \; \rightarrow \; t \bar{t}, \qquad q\bar{q}\; \rightarrow \; t \bar{t}.
\eea
The single top production processes, as, e.g., $qb\to q't$,
$q\bar{q}'\to t\bar{b}$, or $qg\to q't\bar{b}$,
have much smaller cross sections.
Each of the top quarks of (\ref{ggtt}) 
decays into a $b$-quark and a $W$-boson,
and the $W$-bosons decay into a fermion-antifermion pair each which leads to
reactions with 6 fermions in the final state. 
The top quark pair production events are best identified if
one of the $W$ bosons decays leptonically and the other hadronically that
corresponds to reactions of the form
\bea
\label{pp6f}
pp \;\ra\; b q\bar{q}'\;\bar{b} l\bar{\nu}_l.
\eea
This means that one should selects events 
with an isolated electron or muon with large
transverse momentum, a missing transverse momentum from the undetected
neutrino and four or more jets. 

For the sake of clarity, in the present work, we will concentrate on one 
specific channel of (\ref{pp6f}):
\bea
\label{ppbbudmn}
pp \;\ra\; b u\bar{d}\;\bar{b} \mu^-\bar{\nu}_{\mu}
\eea
and address the question to which extent the anomalous $Wtb$ coupling
affects different distributions 
of the final state $\mu^-$. In other words, we would
like to check if the decoupling theorem of \cite{GHR}
holds also for the top quark pair 
production in proton-proton collisions at the LHC.
We will also illustrate the role of the off-resonance background
contributions in (\ref{ppbbudmn}) by comparing
the distributions computed with the full set of leading order Feynman
diagrams with those computed with the $t\bar t$ production 
diagrams only.

\section{An anomalous $Wtb$ coupling at the LHC}

\begin{figure}[htb]
\centerline{
\epsfig{file=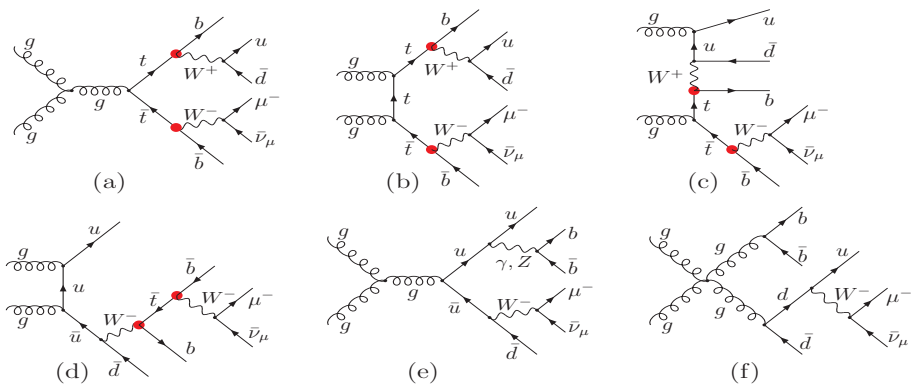,  width=120mm, height=50mm}}
\caption{Examples of the leading order Feynman diagrams of process
(\ref{ggbbudmn}). Blobs indicate the $Wtb$ coupling.}
\label{diags_gg}
\end{figure}

\begin{figure}[htb]
\centerline{
\epsfig{file=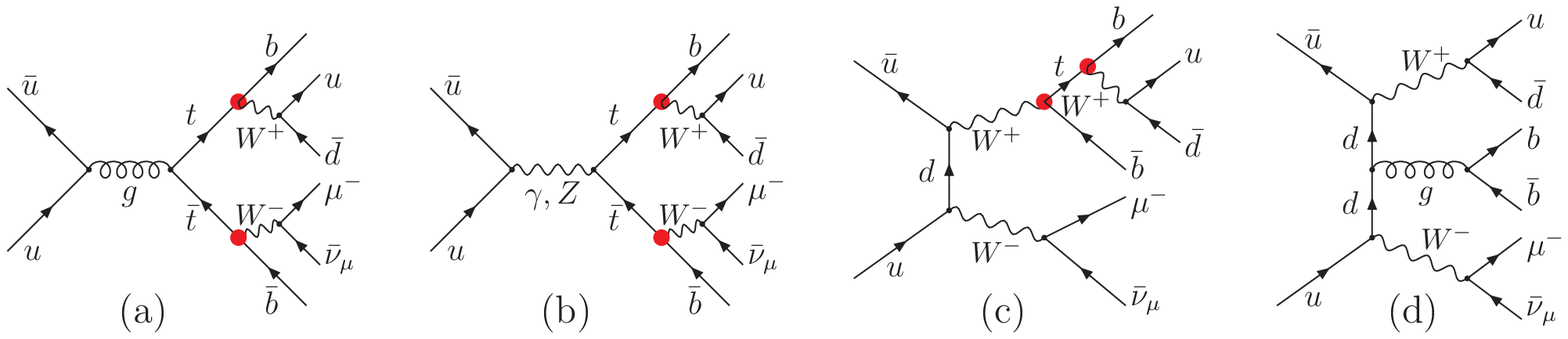,  width=120mm, height=25mm}}
\caption{Examples of the leading order Feynman diagrams of process
(\ref{uubbudmn}). Blobs indicate the $Wtb$ coupling.}
\label{diags_uu}
\end{figure}

The underlying hard scattering processes of (\ref{ppbbudmn}) that contribute
most to its cross section are the following:
\bea
\label{ggbbudmn}
gg &\ra& b u\bar{d}\;\bar{b} \mu^-\bar{\nu}_{\mu},\\
\label{uubbudmn}
u\bar u &\ra& b u\bar{d}\;\bar{b} \mu^-\bar{\nu}_{\mu},\\
\label{ddbbudmn}
d\bar d &\ra& b u\bar{d}\;\bar{b} \mu^-\bar{\nu}_{\mu}.
\eea
In the leading order, neglecting light fermion masses,
$m_u=m_d=m_{\mu}=0$, and the Cabibbo-Kobayashi-Maskawa mixing between quarks, 
there are 421 Feynman diagrams of the gluon-gluon fusion process 
(\ref{ggbbudmn}) and 718 diagrams of each of
the quark-antiquark annihilation processes (\ref{uubbudmn}) and 
(\ref{ddbbudmn}).
Some examples of the diagrams of processes (\ref{ggbbudmn}) and (\ref{uubbudmn})
are shown in Figs.~\ref{diags_gg} and \ref{diags_uu}. There are 3 
double resonance $t\bar t$ production signal diagrams of the gluon-gluon
fusion processes (\ref{ggbbudmn}): two of them are depicted 
in Figs.~\ref{diags_gg}a and \ref{diags_gg}b and
the third is obtained by interchanging the gluon lines in Fig.~\ref{diags_gg}b.
At the same time the quark-antiquark annihilation process
(\ref{uubbudmn}) receives contributions from 6
$t\bar t$ production signal diagrams: 3 of them are depicted in 
Figs.~\ref{diags_uu}a and \ref{diags_uu}b and the other 3 are obtained
by interchanging the $u$-quark lines in each diagram of Figs.~\ref{diags_uu}a 
and \ref{diags_uu}b. The Feynman diagrams of process (\ref{ddbbudmn}) are
obtained from those of process (\ref{uubbudmn}) just by replacing the initial 
state $u$-quarks with $d$-quarks. Note that if the top quarks were assumed 
to be produced
on-shell, which corresponds to the NWA for the top quarks, then the number 
of the signal diagrams for each 
of the processes (\ref{ggbbudmn}), (\ref{uubbudmn}) and (\ref{ddbbudmn}) 
would be equal 3.
Thus, in the NWA, not only does one neglect a plethora of background
contributions represented by the diagrams in Figs.~\ref{diags_gg}e,
\ref{diags_gg}f and \ref{diags_uu}d, but
also part of the double resonance signal diagrams and the diagrams with two
or one top quark propagators, as those
depicted in Figs.~\ref{diags_gg}c, \ref{diags_gg}d and \ref{diags_uu}c.
Let us note that the $Wtb$ coupling that is indicated by a blob enters
twice both in the $t\bar t$ production signal diagrams of 
Figs.~\ref{diags_gg}a, \ref{diags_gg}b, ~\ref{diags_uu}a, \ref{diags_uu}b 
and in the non double resonance diagrams of Figs.~\ref{diags_gg}c, 
\ref{diags_gg}d and \ref{diags_uu}c.

The effective Lagrangian of the 
$Wtb$ interaction containing operators of dimension up to five
considered in the present work has the following form \cite{kane}:
\bea
\label{lagr}
L_{Wtb}&=&\frac{g}{\sqrt{2}}\,V_{tb}\left[W^-_{\mu}\bar{b}\,\gamma^{\mu}
\left(f_1^L P_L +f_1^R P_R\right)t\right.\nn\\
&&\qquad\qquad\qquad
 \left.-\frac{1}{m_W}\partial_{\nu}W^-_{\mu}\bar{b}\,\sigma^{\mu\nu}
  \left(f_2^L P_L +f_2^R P_R\right)t\right]\nn\\
&+&\frac{g}{\sqrt{2}}\,V_{tb}^*\left[W^+_{\mu}\bar{t}\,\gamma^{\mu}
\left(\bar{f}_1^L P_L +\bar{f}_1^R P_R\right)b\right.\nn\\ 
&&\qquad\qquad\qquad \left.
-\frac{1}{m_W}\partial_{\nu}W^+_{\mu}\bar{t}\,\sigma^{\mu\nu}
  \left(\bar{f}_2^L P_L +\bar{f}_2^R P_R\right)b\right],
\eea
where form factors $f_{i}^{L}$, $f_{i}^{R}$, $\bar{f}_{i}^{L}$ and 
$\bar{f}_{i}^{R}$, $i=1,2$ can be complex in general and 
the remaining notation is obvious. See \cite{afbKK} for details
and the corresponding Feynman rules.
The lowest order SM Lagrangian of $Wtb$ interaction is reproduced 
for $f_{1}^{L}=\bar{f}_{1}^{L}=1$ and all the other form factors equal 0.
CP conservation leads to the following relationships between
the form factors of (\ref{lagr}):
\beq
\label{rel}
\left.\bar{f}_1^{R}\right.^*=f_1^R, 
\quad \left.\bar{f}_1^{L}\right.^*=f_1^L, \qquad \qquad
\left.\bar{f}_2^R\right.^*=f_2^L, \quad \left.\bar{f}_2^L\right.^*=f_2^R.
\eeq

First direct limits on the form factors of (\ref{lagr}) were
obtained by the CDF Collaboration by investigating two form 
factors at a time while assuming the other two at their SM values
\cite{wtbCDF}. The limits
have been improved recently by the D0 Collaboration \cite{wtbD0} 
and they read:
\bea
\label{limits_2dim}
\left|V_{tb}f_{1}^R\right|^2 < 0.93, \qquad
\left|V_{tb}f_{2}^R\right|^2 < 0.13, \qquad
\left|V_{tb}f_{2}^L\right|^2 < 0.06.
\eea
More stringent are one-dimensional limits at 95\% C.L. \cite{wtbD0}:
\bea
\label{limits_1dim}
\left|V_{tb}f_{1}^R\right|^2 < 0.50, \qquad
\left|V_{tb}f_{2}^R\right|^2 < 0.11, \qquad
\left|V_{tb}f_{2}^L\right|^2 < 0.05.
\eea
Limits derived from ATLAS \cite{wtbATLAS} and CMS \cite{wtbCMS} measurements 
of $W$-boson helicity fractions using a program TOPFIT \cite{wtbAguilar}
with one non-zero coupling at a time 
and $V_{tb}= 1$ read:
\bea
\label{limitsLHC}
{\rm Re}f_{1}^R\in  [-0.20,0.23], \quad
{\rm Re}f_{2}^R\in  [-0.08,0.04],\quad
{\rm Re}f_{2}^L\in  [-0.14,0.11].
\eea
Limits (\ref{limitsLHC}) are weaker if two couplings are varied 
at a time \cite{wtbCMS}. Amazingly enough, the right-handed vector form factor
$f_{1}^R$ is least constraint in (\ref{limits_2dim})--(\ref{limitsLHC}), but 
if CP is conserved then it
is indirectly constrained from the CLEO data on
$b\rightarrow s\gamma$ \cite{cleo} and from other rare $B$ 
decays \cite{fajfer}. 

\section{Results}

Lagrangian (\ref{lagr}) was implemented 
into {\tt carlomat} \cite{carlomat}, a general purpose program for the Monte 
Carlo (MC) computation of lowest order cross sections.
A new version of the program \cite{carlomat2} was already used
to make predictions for the FBA in top 
quark production at the Tevatron \cite{afbKK}. 
In this section, a sample of results is presented that illustrate an influence
of the tensor form factors of Lagrangian (\ref{lagr})
on the distributions of the secondary $\mu^-$ 
in top quark pair production at the $pp$ collisions at the LHC energies
through reaction (\ref{ppbbudmn}). As in \cite{afbKK}, we put
$V_{tb}=1$ and assume form factors $f_{i}^{L}$, $f_{i}^{R}$, $\bar{f}_{i}^{L}$ and 
$\bar{f}_{i}^{R}$, $i=1,2$, of Lagrangian (\ref{lagr}) to be real. 
Moreover, the vector form factors are fixed at their SM values of
$f_{1}^{L}=\bar{f}_{1}^{L}=1$, $f_{1}^{R}=\bar{f}_{1}^{R}=0$ and
only the tensor form factors $f_{2}^{L,R}$, $\bar{f}_{2}^{L,R}$ are changed by
assigning them two values: 0 or 0.2 in different, CP-even or CP-odd,
combinations. A value of 0.2 that exceeds limits 
(\ref{limits_2dim})--(\ref{limitsLHC}) is
chosen just for the sake of illustration of the anomalous coupling effects
that for smaller values of the form factors would have been hardly
visible in the plots. However, an interested user can easily obtain
predictions for any other choice of the form factors with the
publicly available program \cite{carlomat2}.

The physical input parameters that are 
used in the computation are the same as in \cite{afbKK}. In order to avoid 
on-shell poles the following complex mass parameters: 
\beq
\label{m2}
m_b^2 \ra M_b^2=m_b^2-im_b\Gamma_b, \;\;\; b=Z, W, h,\quad
m_t\ra M_t=\sqrt{m_t^2-im_t\Gamma_t}
\eeq
are used instead of masses in propagators of unstable particles,
both in the $s$- and $t$-channel.
The particle widths in (\ref{m2}) are assumed to be constant and the square 
root with positive real part is chosen, see \cite{carlomat} for details. 
The computation is performed in the complex mass scheme, where the electroweak 
(EW) couplings are parametrized in terms
the complex EW mixing parameter $\sin^2\theta_W=1-{M_W^2}/{M_Z^2}$
which preserves the lowest order Ward identities \cite{Racoon} and minimizes
the unitarity violation effects at high energies.

The width of the top quark has substantial influence on the cross 
section of reaction (\ref{ppbbudmn}). Therefore, it is calculated 
in the leading order for every specific choice of the form factors 
with effective Lagrangian (\ref{lagr}) that, due to the fact that branching 
fractions of the decays $t\to bW^+$ and $\bar t\to \bar{b}W^-$ are close to 1,
practically determines it.
It should be stressed that for CP-odd choices of the form factors the widths
$\Gamma_t$ of $t$ and $\Gamma_{\bar t}$ of $\bar t$ differ from each other.
Thus, the both widths are calculated and the following rule is applied
to put the width in the $s$-channel top quark propagator: 
$\Gamma_t$ is used if the propagator goes into $bW^+$ and 
$\Gamma_{\bar t}$ is used if the propagator goes into $\bar{b}W^-$. The rule does
not work for the propagators in $t$-channel, but the actual value
of the top quark width does not play much of a role there.
If the prescription is applied then the $t\bar t$ production signal 
contribution to the cross section of (\ref{ppbbudmn}) in the NWA takes the 
following form
\bea
&&\left.\sigma(pp\to t^*\bar t^* \ra b u\bar{d}\;\bar{b} \mu^-\bar{\nu}_{\mu})
\right|_{\rm NWA}\nn\\
&&\qquad\qquad\qquad
\approx\sigma(pp\to t\bar t)\; \frac{\Gamma_{t\to bW^+}}{\Gamma_t}\;
\frac{\Gamma_{W^+\to u\bar{d}}}{\Gamma_{W}}\;
\frac{\Gamma_{\bar t\to \bar{b}W^-}}{\Gamma_{\bar t}}\;
\frac{\Gamma_{W^-\to \mu^-\bar{\nu}_{\mu}}}{\Gamma_{W}}\nn\\
&&\qquad\qquad\qquad
\approx\sigma(pp\to t\bar t)\; 
\frac{\Gamma_{W^+\to u\bar{d}}}{\Gamma_{W}}\;
\frac{\Gamma_{W^-\to \mu^-\bar{\nu}_{\mu}}}{\Gamma_{W}}\nn
\eea
and unitarity is preserved. The unitarity argument can be also used to justify 
the prescription for the $s$-channel top quark propagator in the off resonance 
background Feynman diagrams which, as will be illustrated later, contribute 
very little to the cross section.
However, a field theoretical
justification of the effective prescription proposed would actually
require calculation of higher order corrections to partonic
processes (\ref{ggbbudmn})--(\ref{ddbbudmn})
with nonrenormalizable Lagrangian (\ref{lagr}). This formidable and delicate
task is beyond the scope of this work.

The cross section of reaction (\ref{ppbbudmn}) 
is calculated by folding CTEQ6L parton distribution 
functions (PDFs) \cite{CTEQ} with the cross sections of underlying
hard scattering processes
(\ref{ggbbudmn}), (\ref{uubbudmn}) and (\ref{ddbbudmn}), with the cross
sections of the quark-antiquark annihilation processes being symmetrized
with respect to the interchange of the initial state quark and antiquark.
The factorization scale is assumed to be equal 
$Q = \sqrt{m_t^2+\sum_jp_{Tj}^2}$.
The $t\bar{t}$ production events are identified with the following 
acceptance cuts on the transverse momenta $p_T$, pseudorapidities $\eta$, 
missing transverse energy $/\!\!\!\!E^T$ and separation 
$\Delta R_{ik}=\sqrt{\left(\eta_i-\eta_k\right)^2
+\left(\varphi_i-\varphi_k\right)^2}$ in the 
pseudorapidity--azimuthal angle $(\varphi)$ plane between 
the objects $i$ and $k$:
$$p_{Tl} > 30\;{\rm GeV}/c, \qquad p_{Tj} > 30\;{\rm GeV}/c, \qquad
\left|\eta_l\right| < 2.1, \qquad \left|\eta_j\right| < 2.4,$$
\bea
\label{cuts}
/\!\!\!\!E^T > 20\;{\rm GeV},\qquad
\Delta R_{lj,jj} > 0.4.
\eea
The subscripts $l$ and $j$ in (\ref{cuts}) stand for {\em lepton} 
and {\em jet}, a direction of the latter is identified with the 
direction of the corresponding quark. Cuts (\ref{cuts}) are rather restrictive
which means that only slightly more than 1\% 
of the MC events generated by {\tt carlomat} pass them.

\begin{figure}[htb]
\vspace{100pt}
\begin{center}
\setlength{\unitlength}{1mm}
\begin{picture}(35,35)(0,0)
\includegraphics{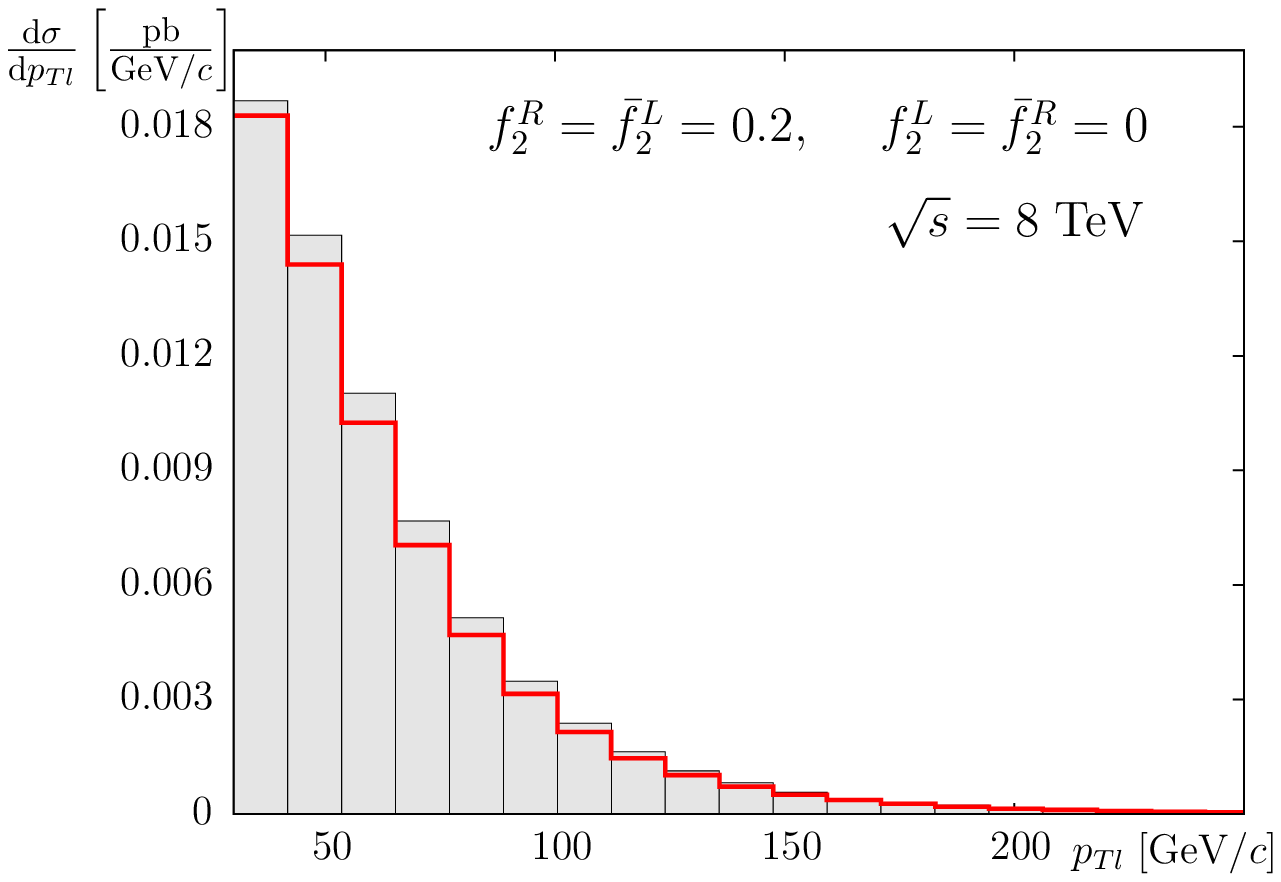}
\end{picture}
\hfill
\begin{picture}(35,35)(0,0)
\includegraphics{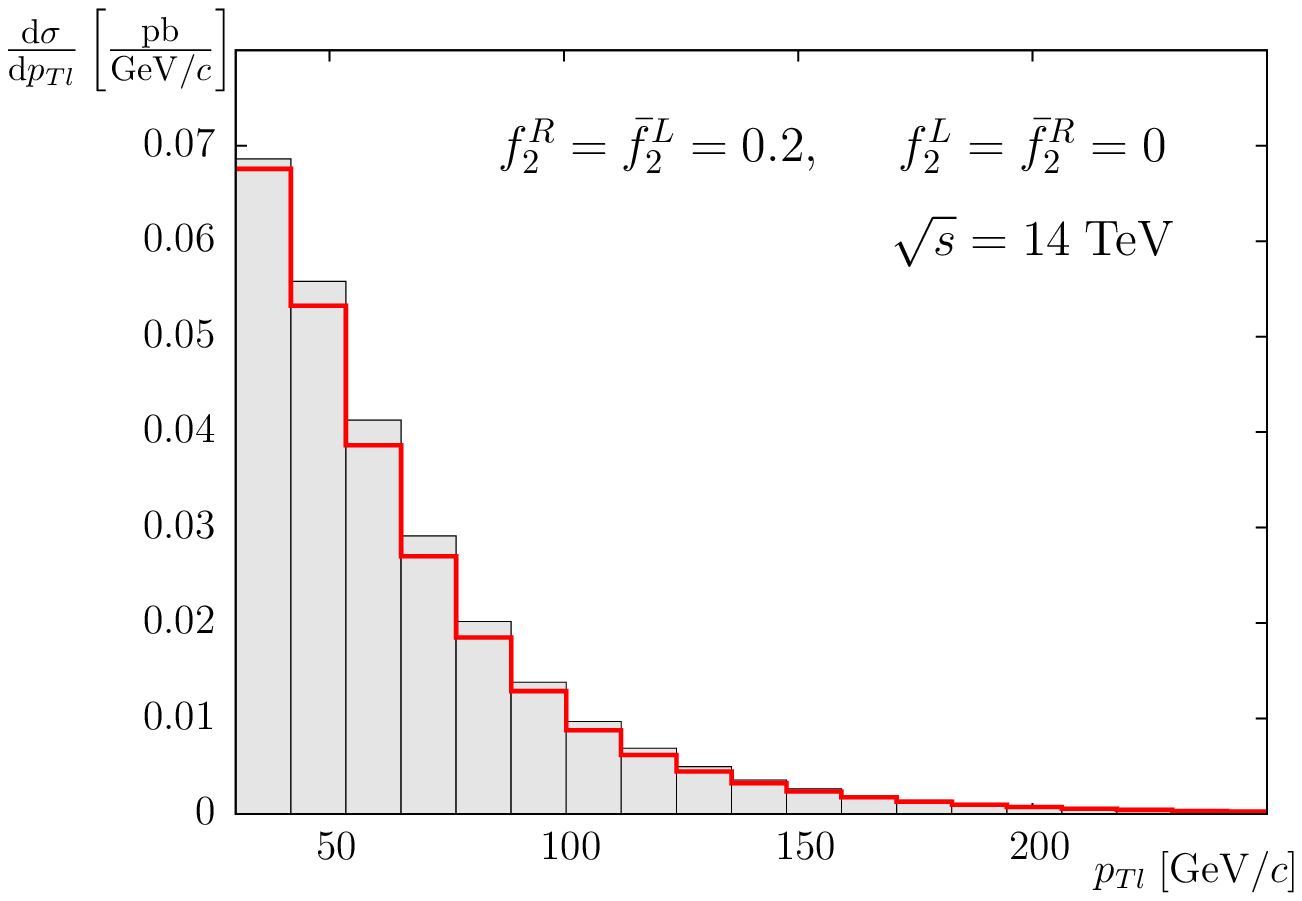}
\end{picture}\\[1.cm]
\begin{picture}(35,35)(0,0)
\includegraphics{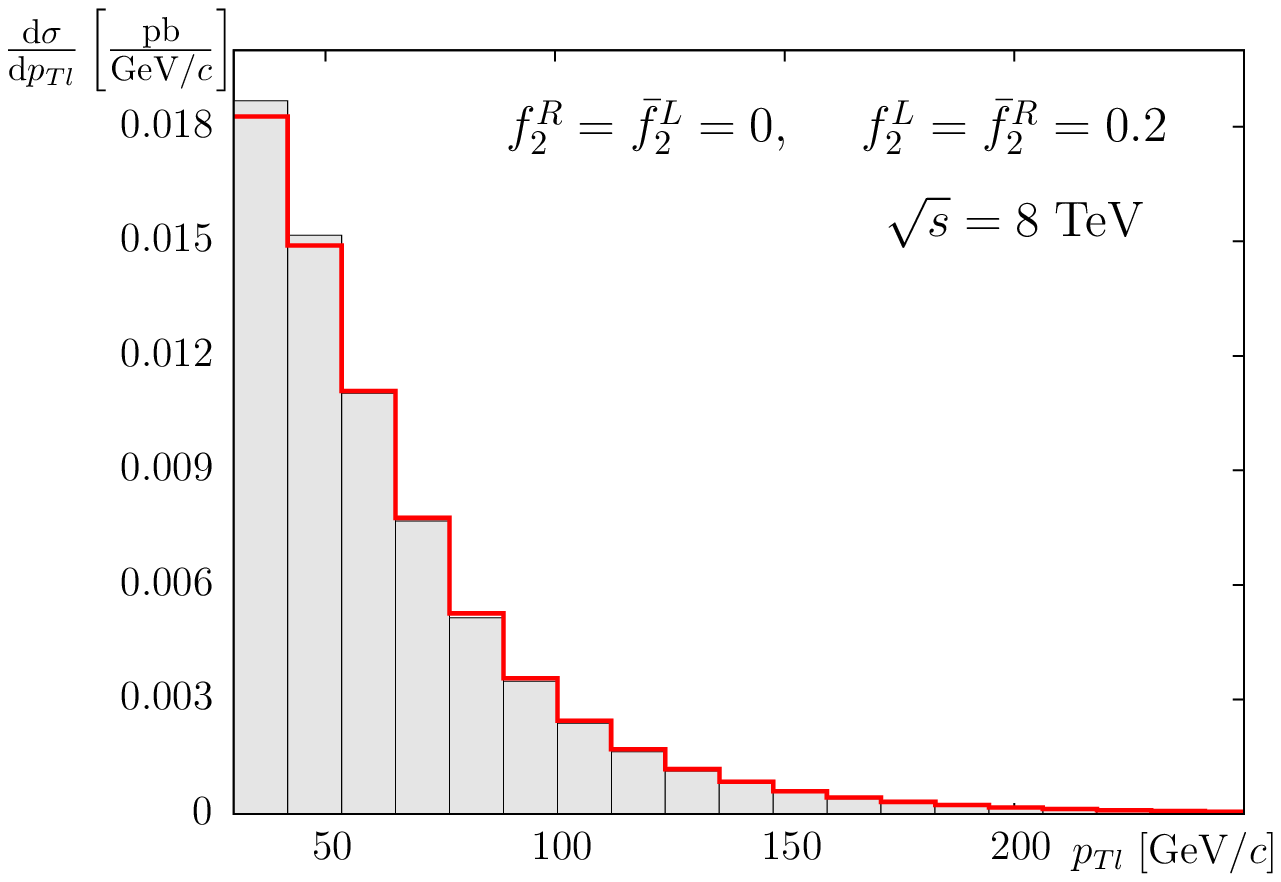}
\end{picture}
\hfill
\begin{picture}(35,35)(0,0)
\includegraphics{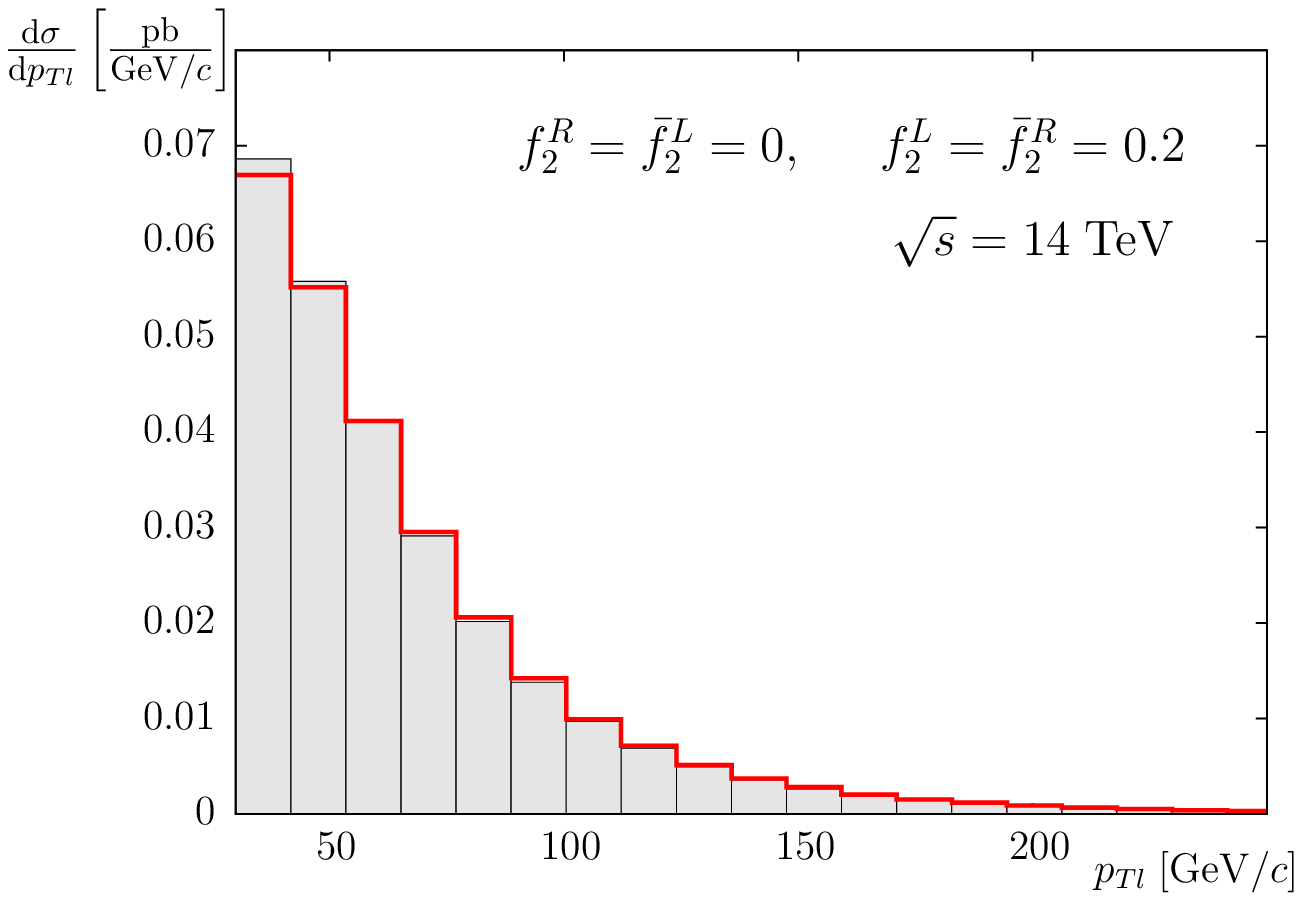}
\end{picture}\\[1.cm]
\begin{picture}(35,35)(0,0)
\includegraphics{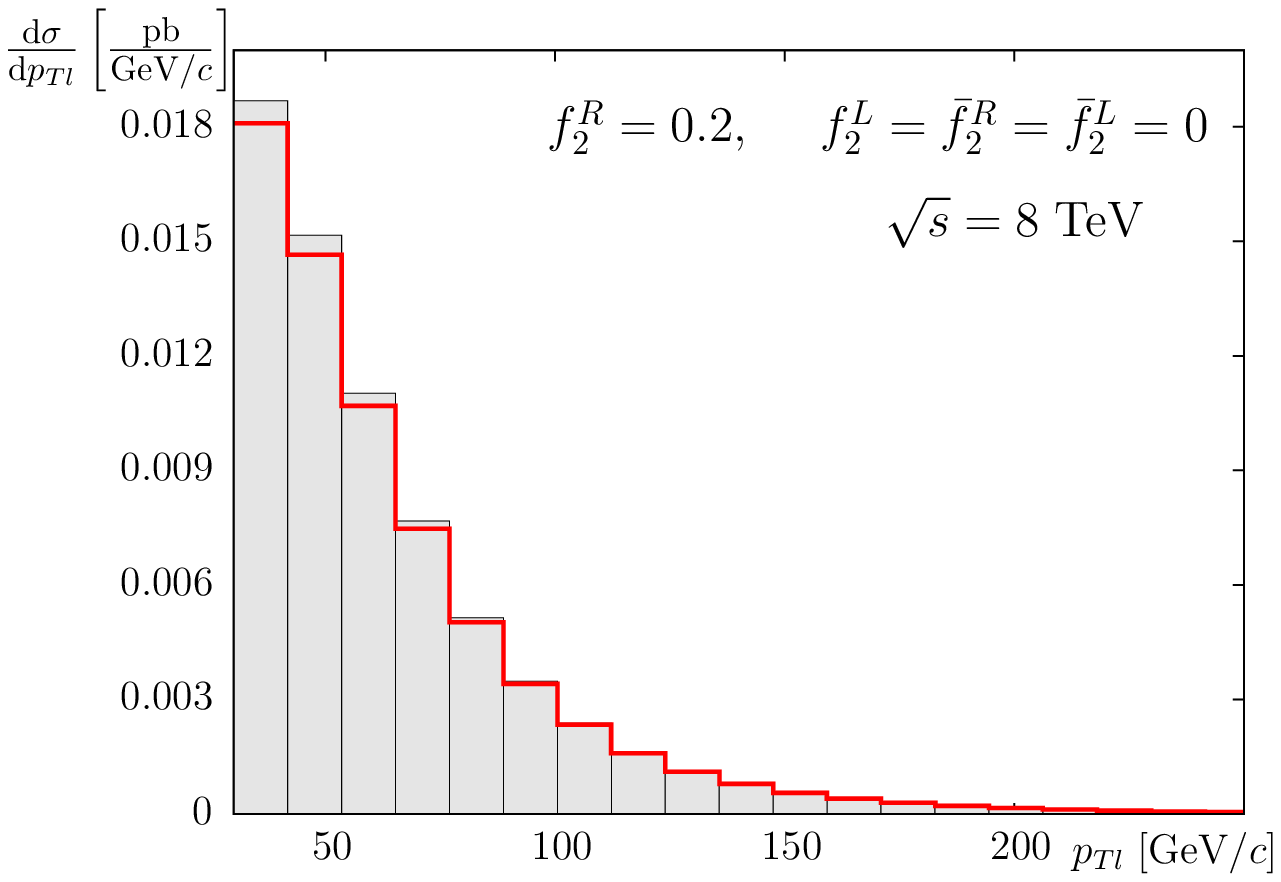}
\end{picture}
\hfill
\begin{picture}(35,35)(0,0)
\includegraphics{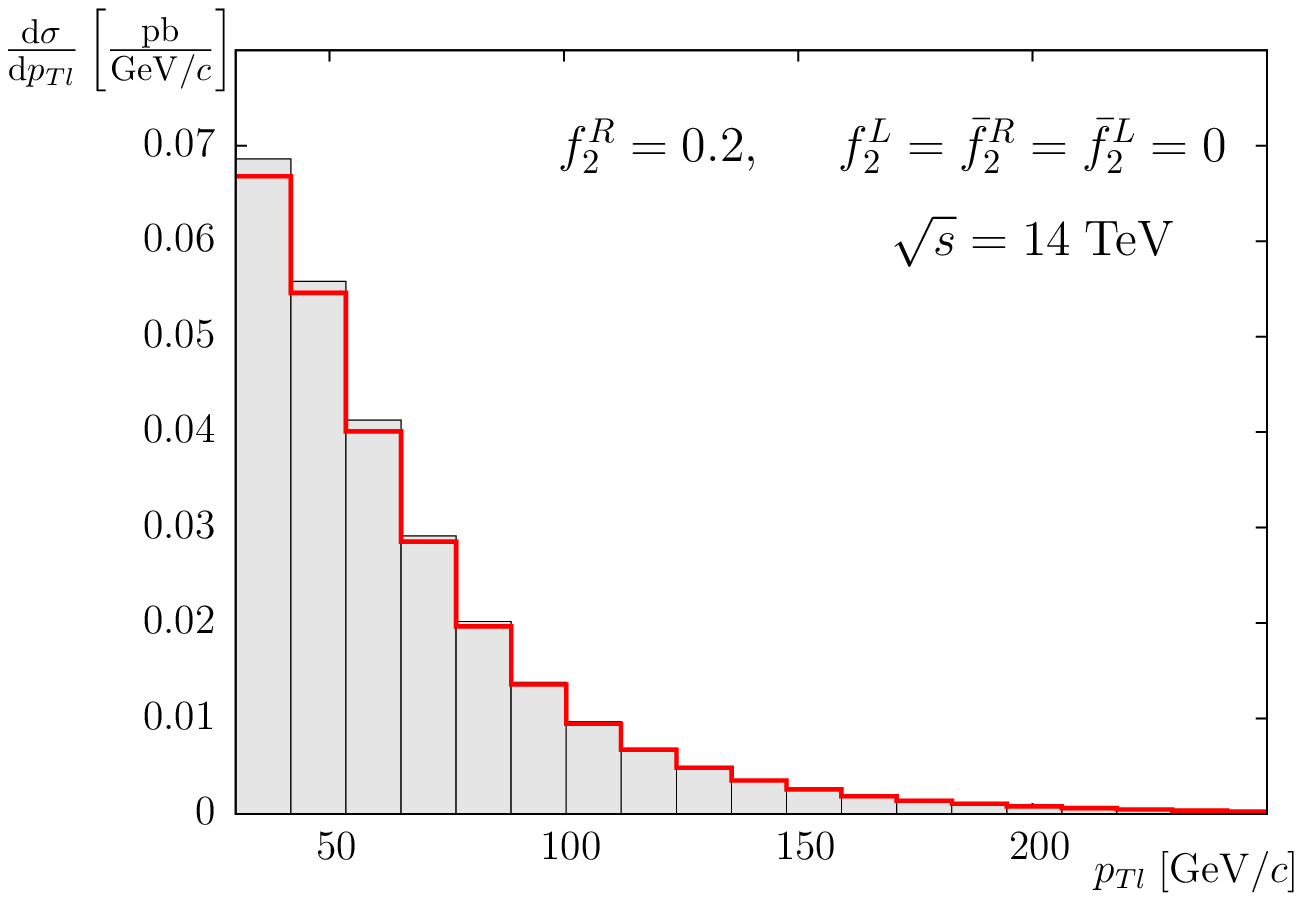}
\end{picture}\\[1.cm]
\begin{picture}(35,35)(0,0)
\includegraphics{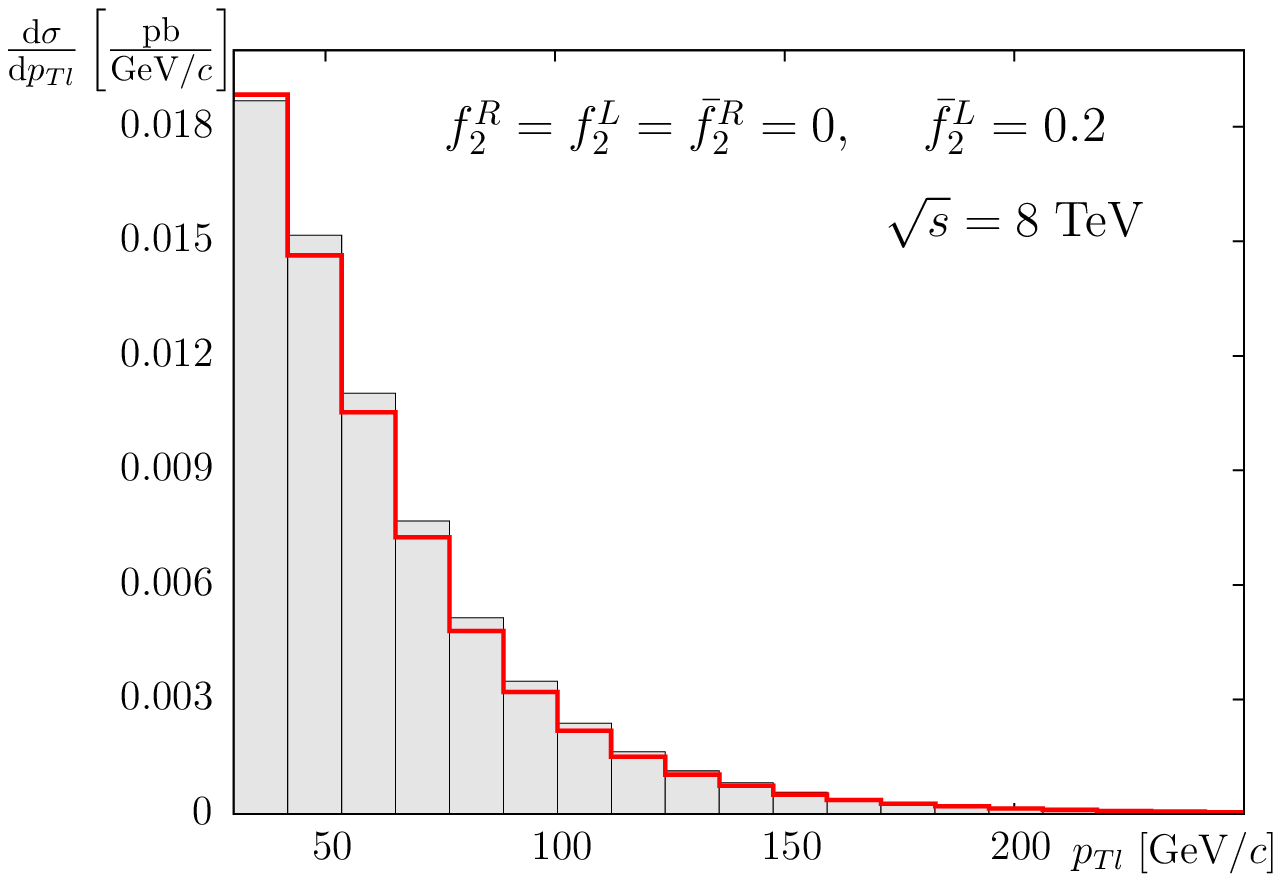}
\end{picture}
\hfill
\begin{picture}(35,35)(0,0)
\includegraphics{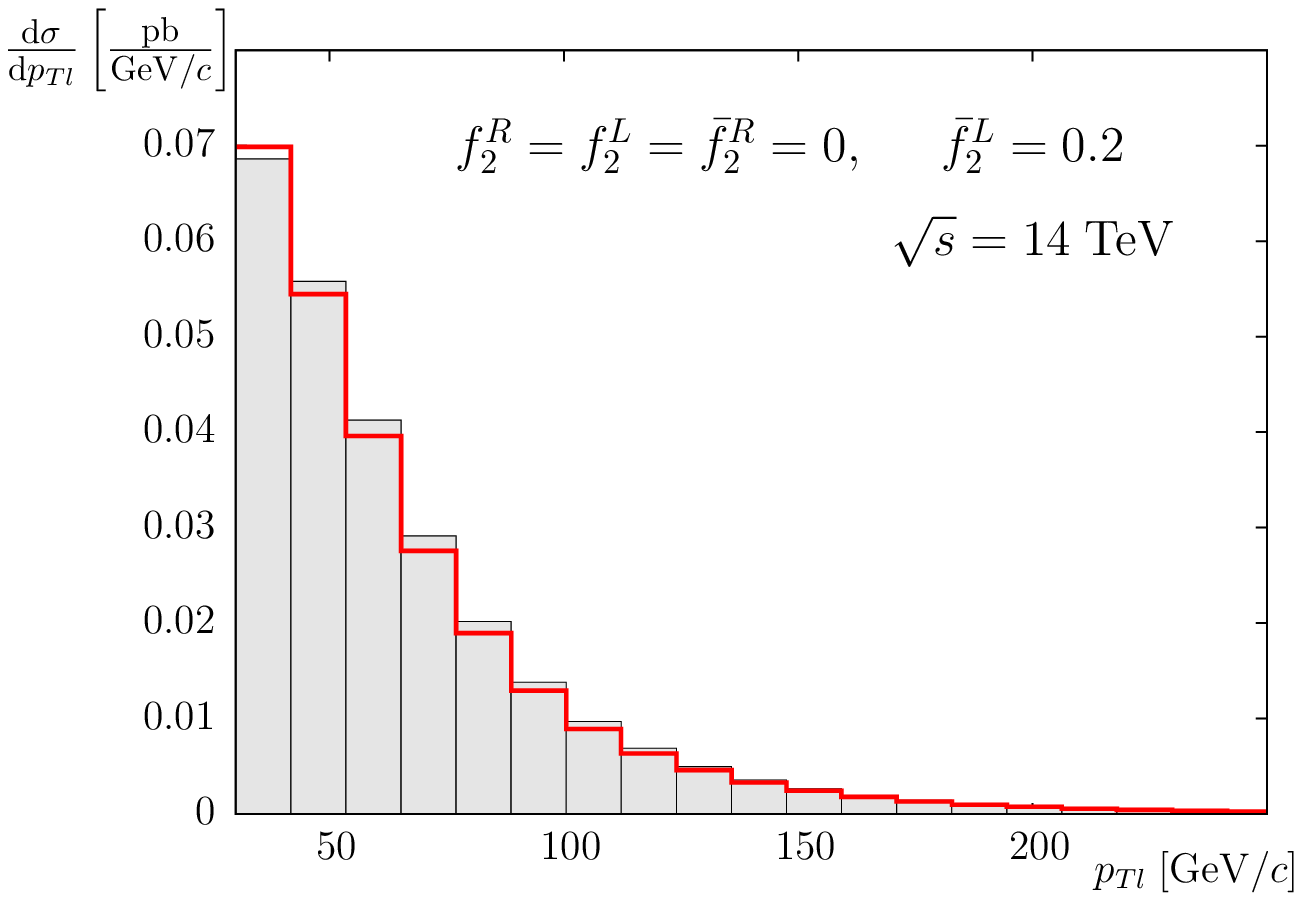}
\end{picture}
\end{center}
\vspace*{-2.cm}
\caption{Distributions in $p_T$ of the final state $\mu^-$ of reaction 
(\ref{ppbbudmn}) in $pp$ collisions at $\sqrt{s}=8$~TeV (left) and 
$\sqrt{s}=14$~TeV 
(right) for CP-even (two upper rows) and CP-odd (two lower rows) 
choices of the tensor form factors of (\ref{lagr}).
}
\label{figptl}
\end{figure}

\begin{figure}[htb]
\vspace{100pt}
\begin{center}
\setlength{\unitlength}{1mm}
\begin{picture}(35,35)(0,0)
\includegraphics{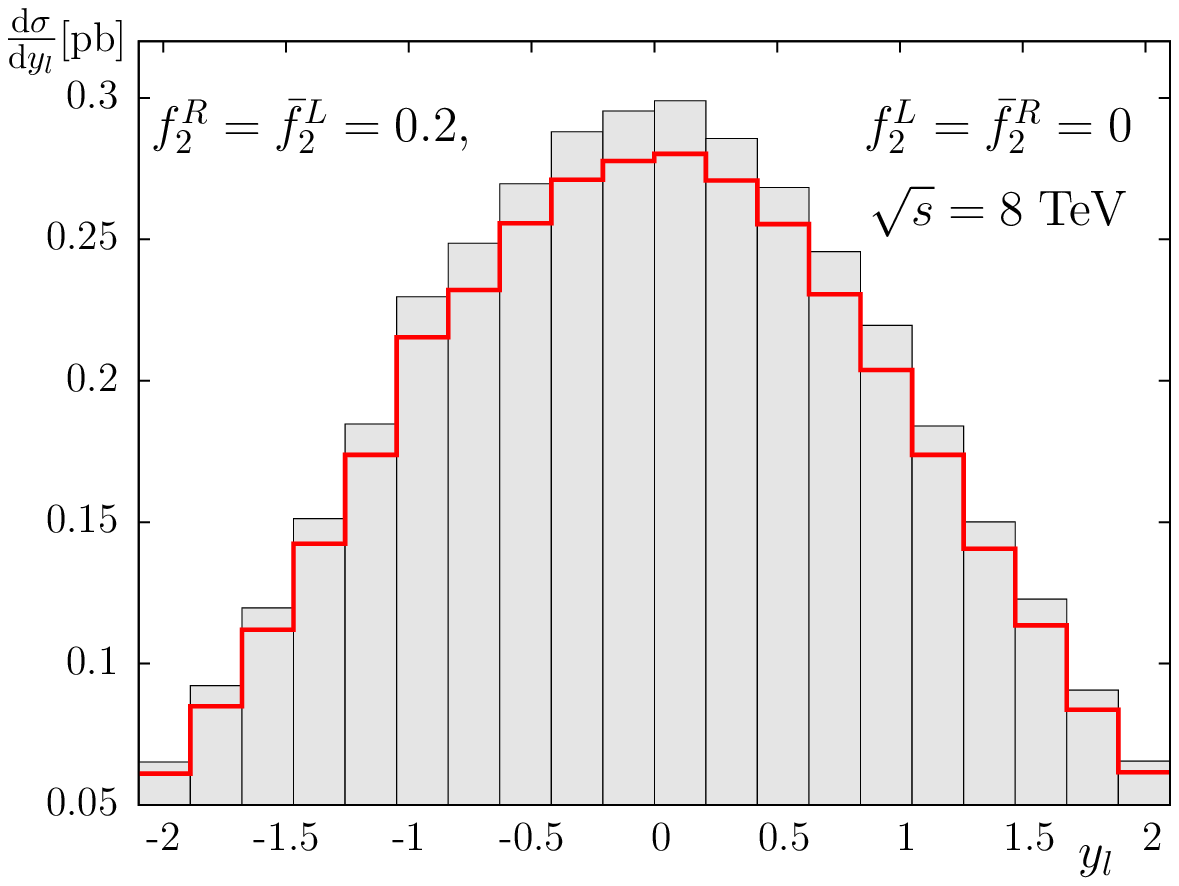}
\end{picture}
\hfill
\begin{picture}(35,35)(0,0)
\includegraphics{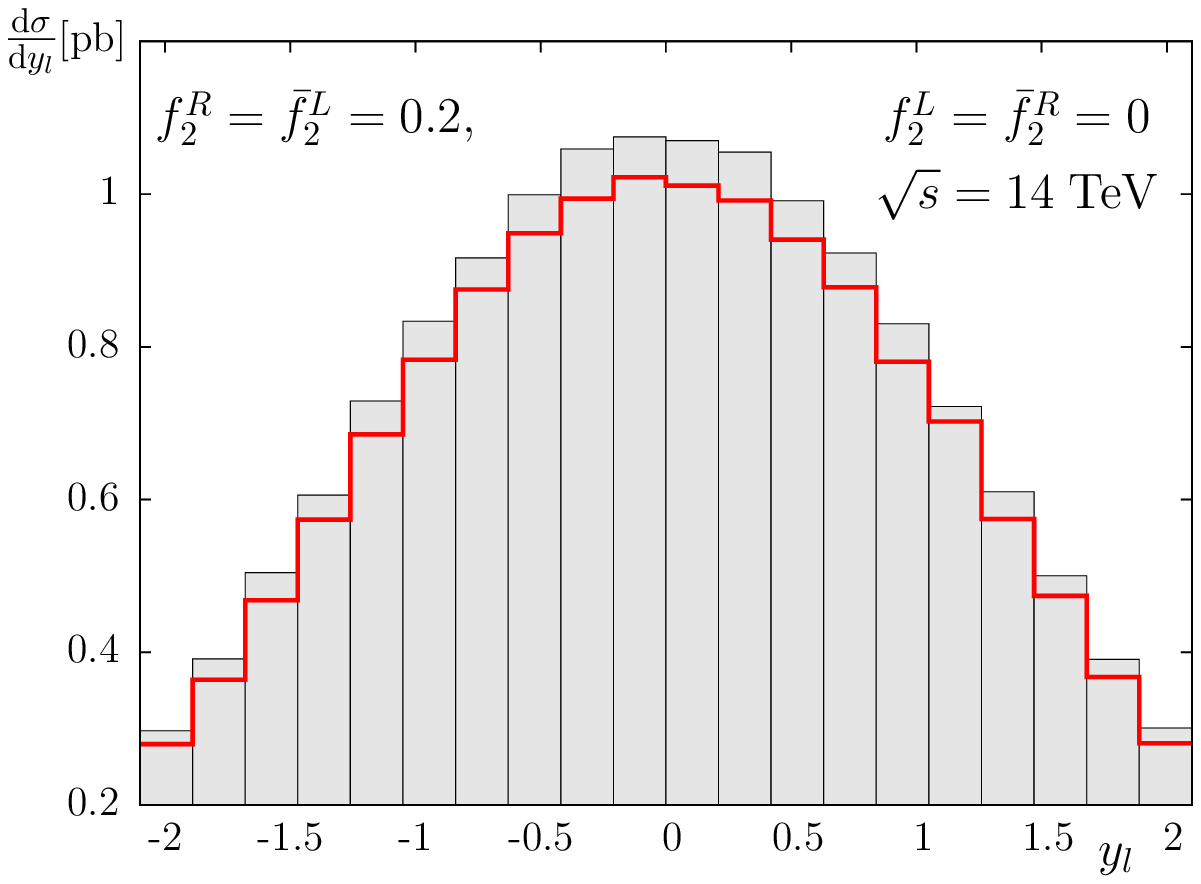}
\end{picture}\\[1.cm]
\begin{picture}(35,35)(0,0)
\includegraphics{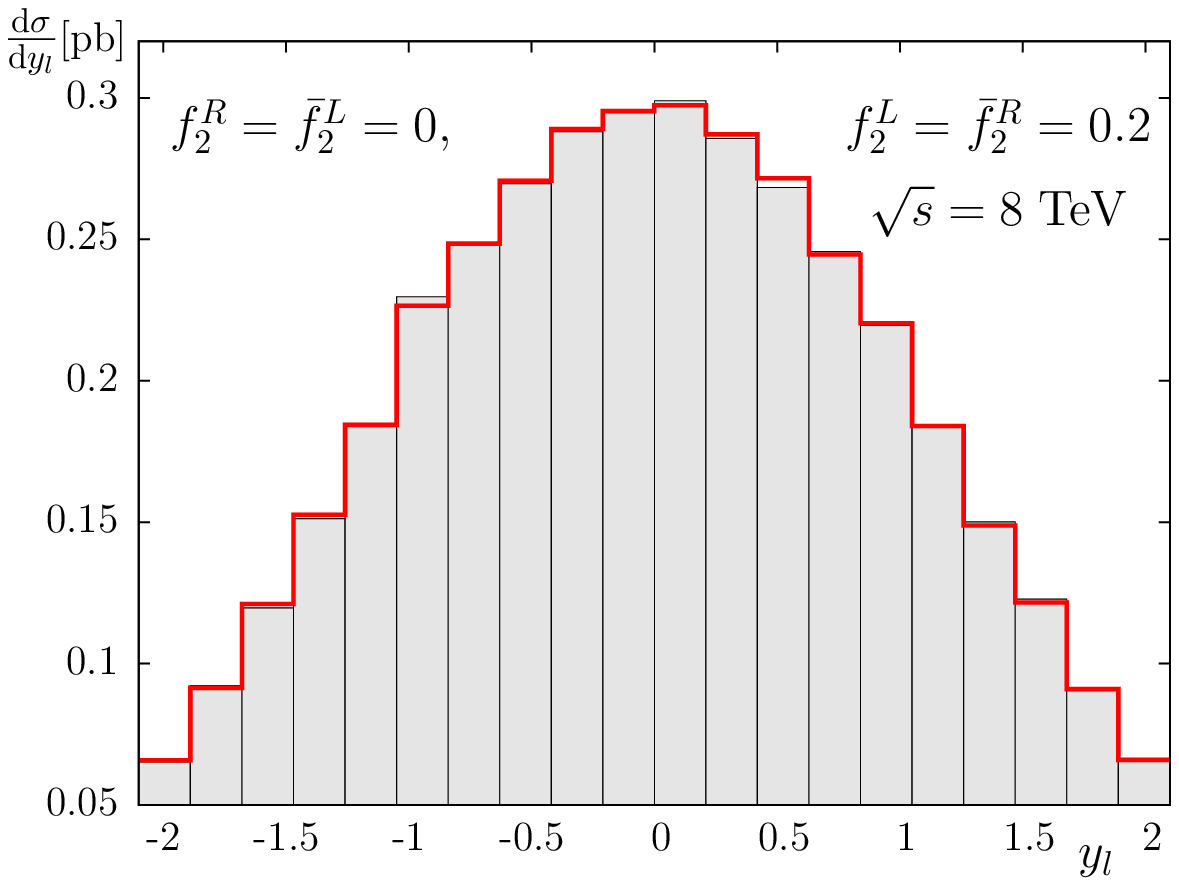}
\end{picture}
\hfill
\begin{picture}(35,35)(0,0)
\includegraphics{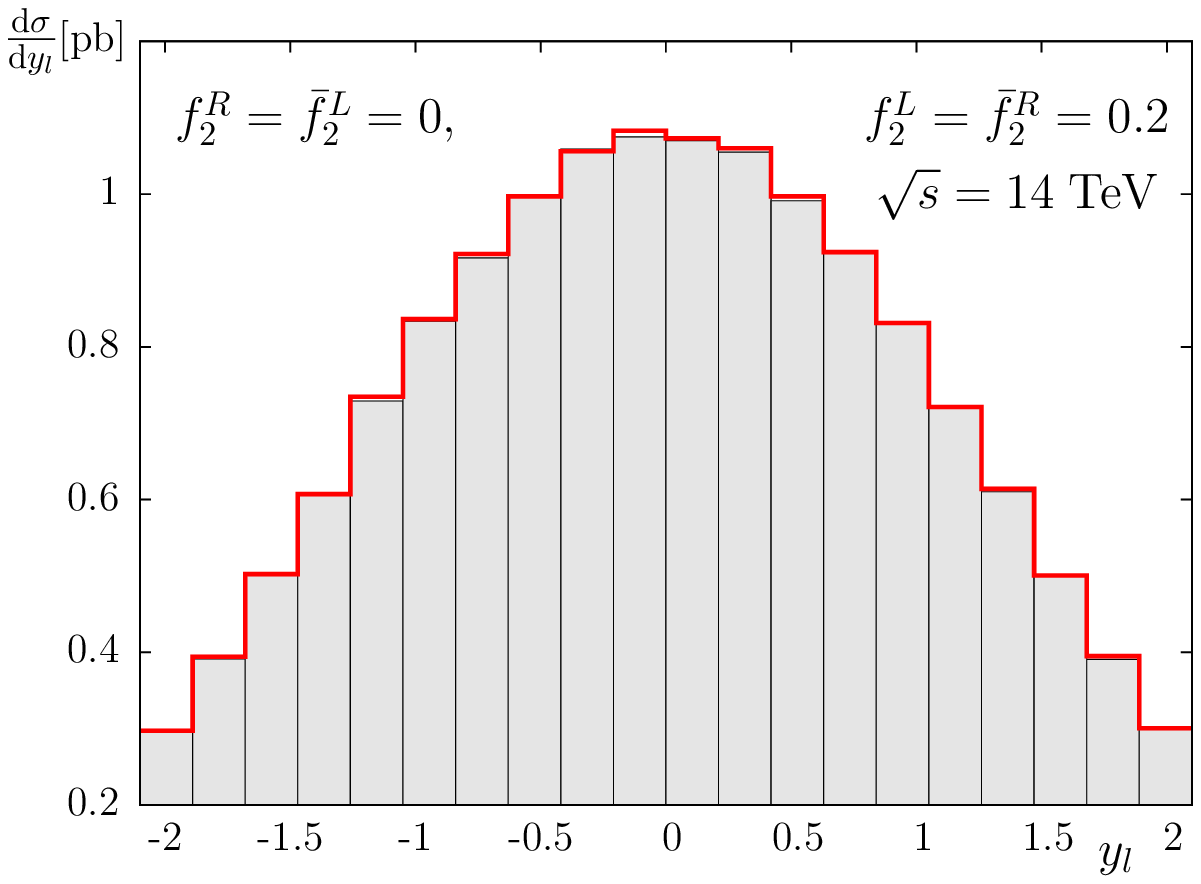}
\end{picture}\\[1.cm]
\begin{picture}(35,35)(0,0)
\includegraphics{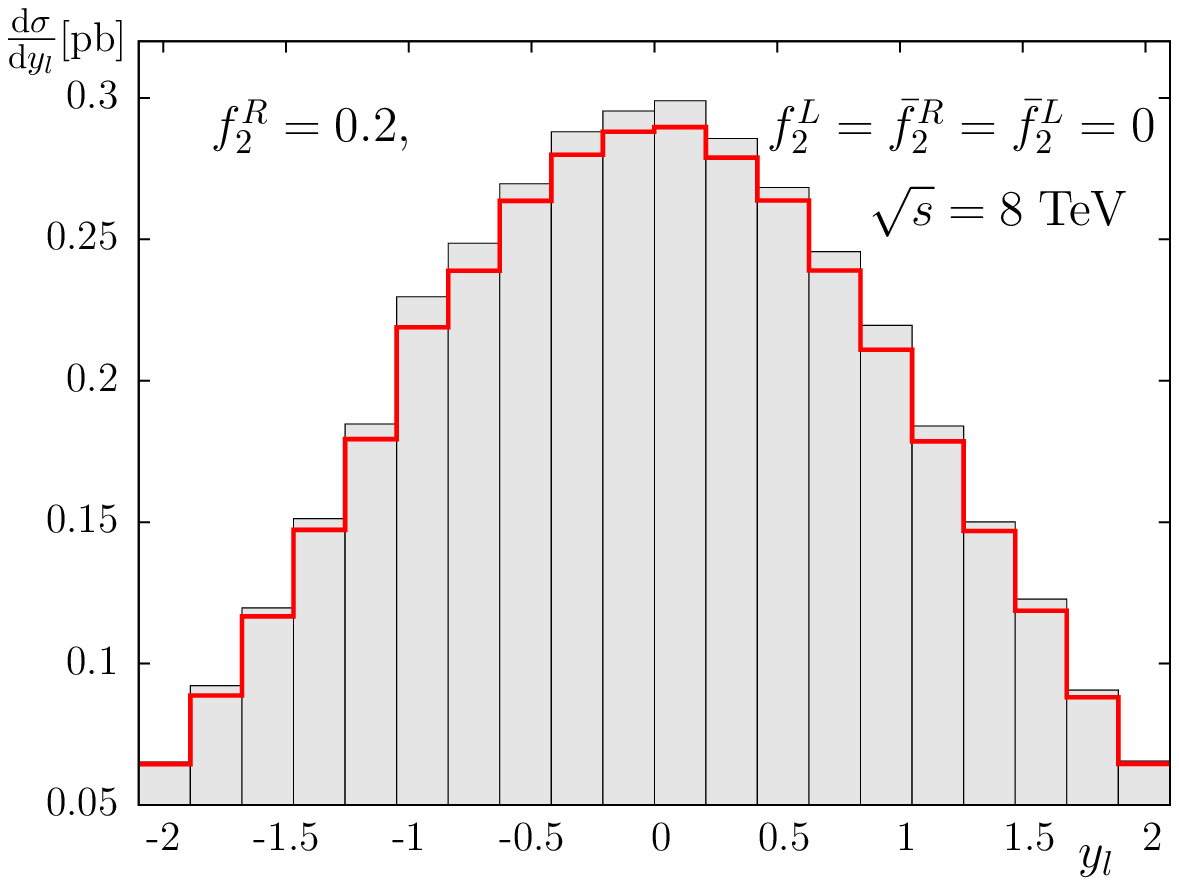}
\end{picture}
\hfill
\begin{picture}(35,35)(0,0)
\includegraphics{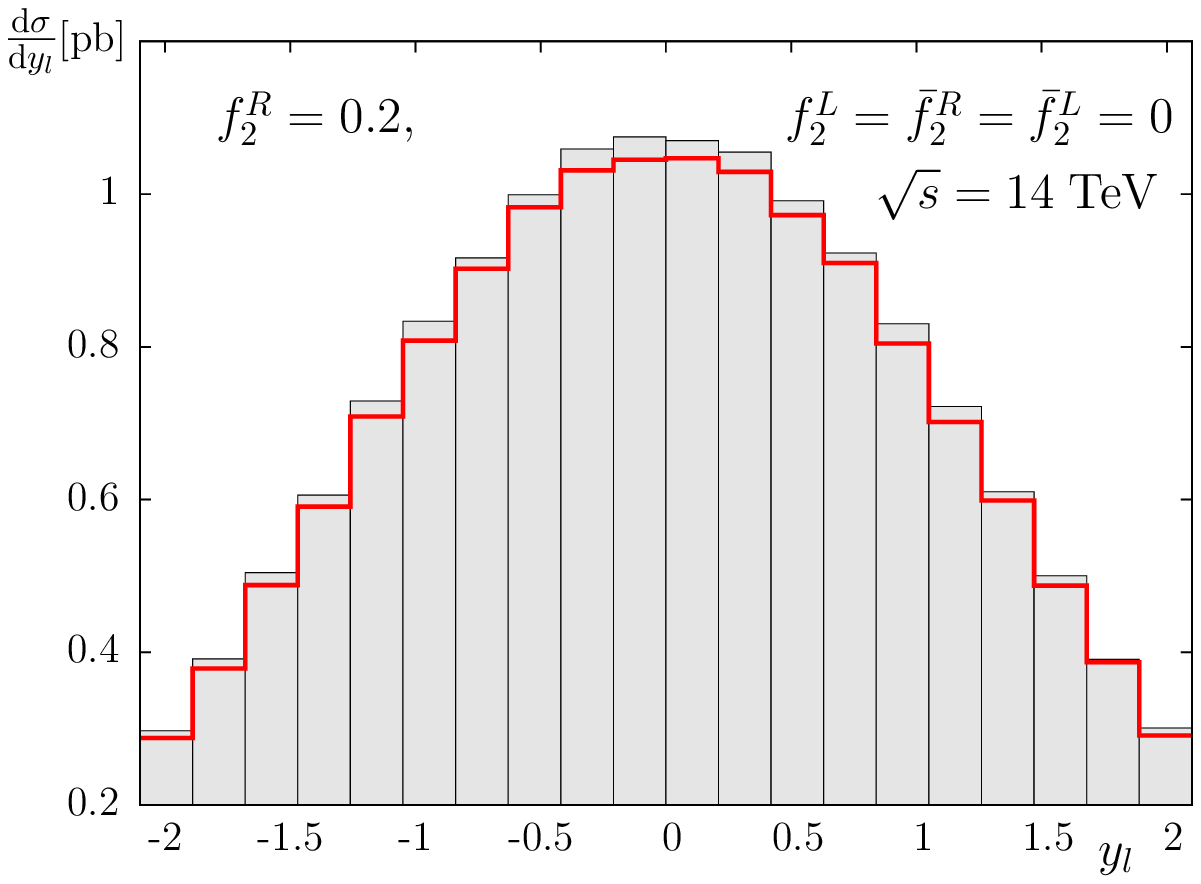}
\end{picture}\\[1.cm]
\begin{picture}(35,35)(0,0)
\includegraphics{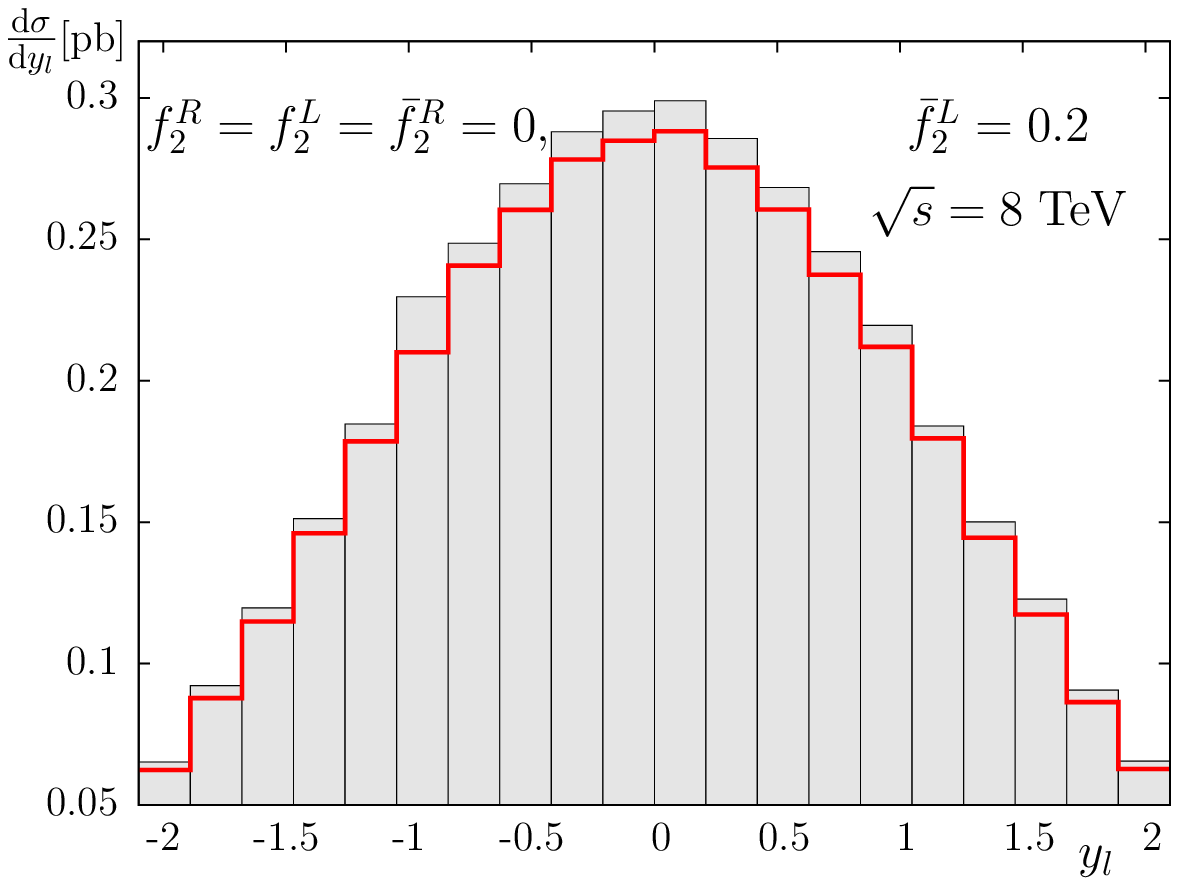}
\end{picture}
\hfill
\begin{picture}(35,35)(0,0)
\includegraphics{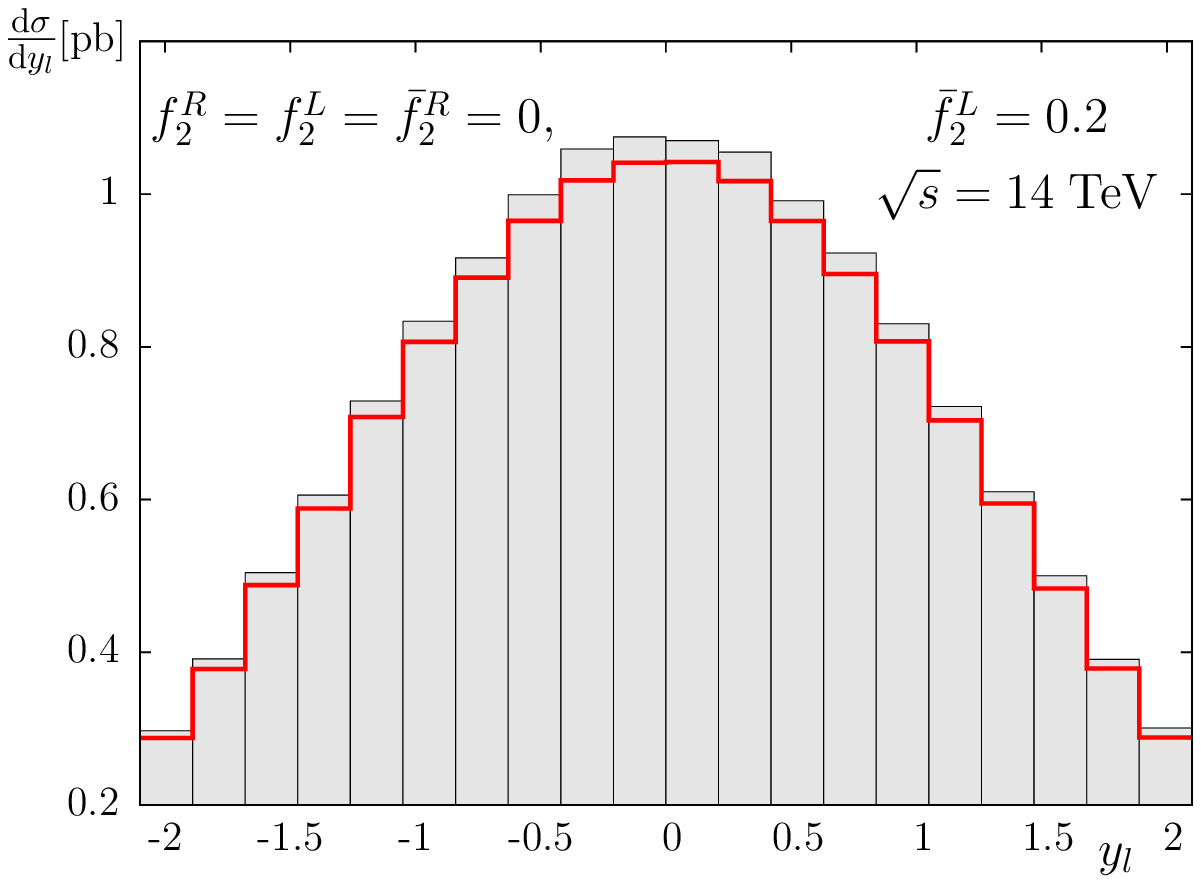}
\end{picture}
\end{center}
\vspace*{-2.cm}
\caption{Distributions in rapidity of the final state $\mu^-$ of reaction 
(\ref{ppbbudmn}) in $pp$ collisions at $\sqrt{s}=8$~TeV (left) and 
$\sqrt{s}=14$~TeV 
(right) for CP-even (two upper rows) and CP-odd (two lower rows) 
choices of the tensor form factors of (\ref{lagr}).}
\label{figrapl}
\end{figure}

\begin{figure}[htb]
\vspace{100pt}
\begin{center}
\setlength{\unitlength}{1mm}
\begin{picture}(35,35)(0,0)
\includegraphics{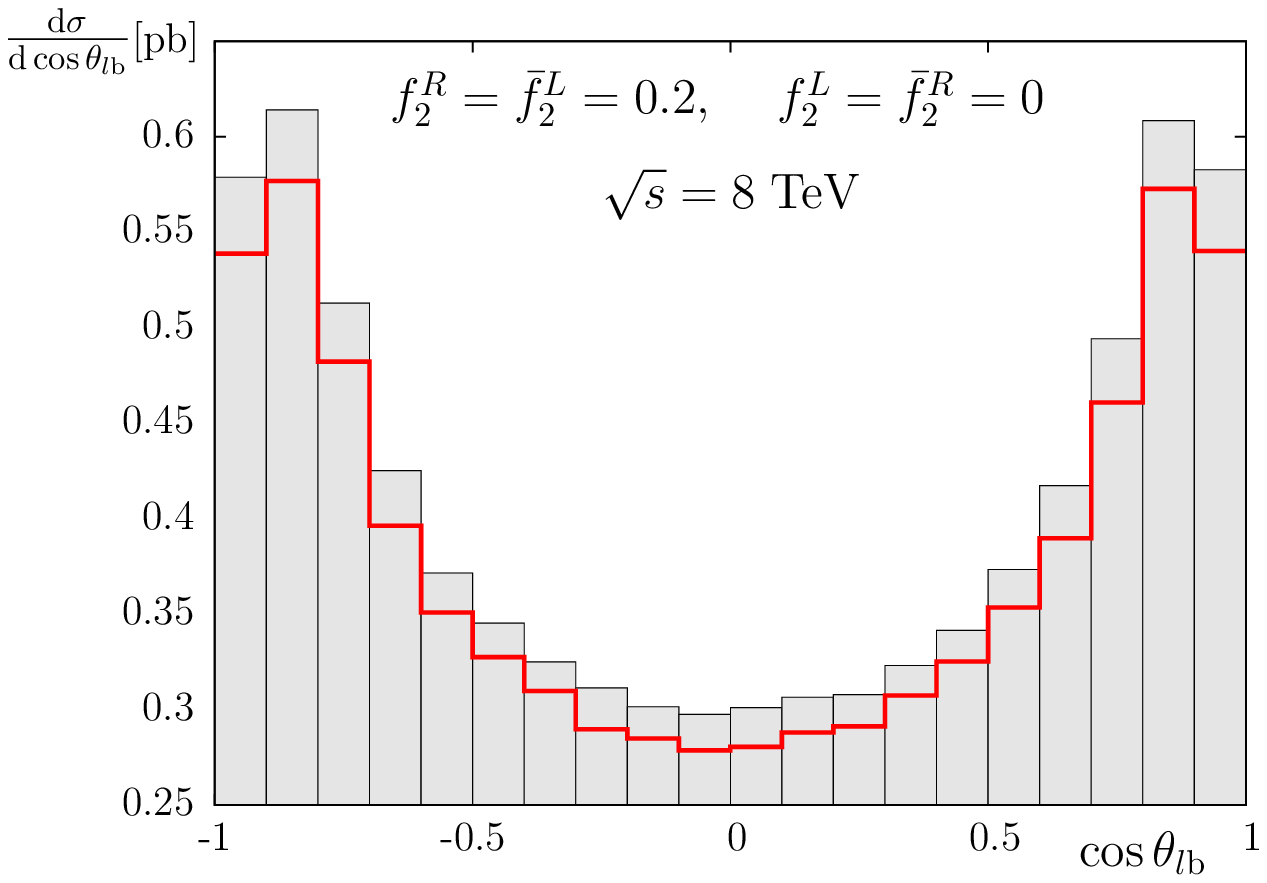}
\end{picture}
\hfill
\begin{picture}(35,35)(0,0)
\includegraphics{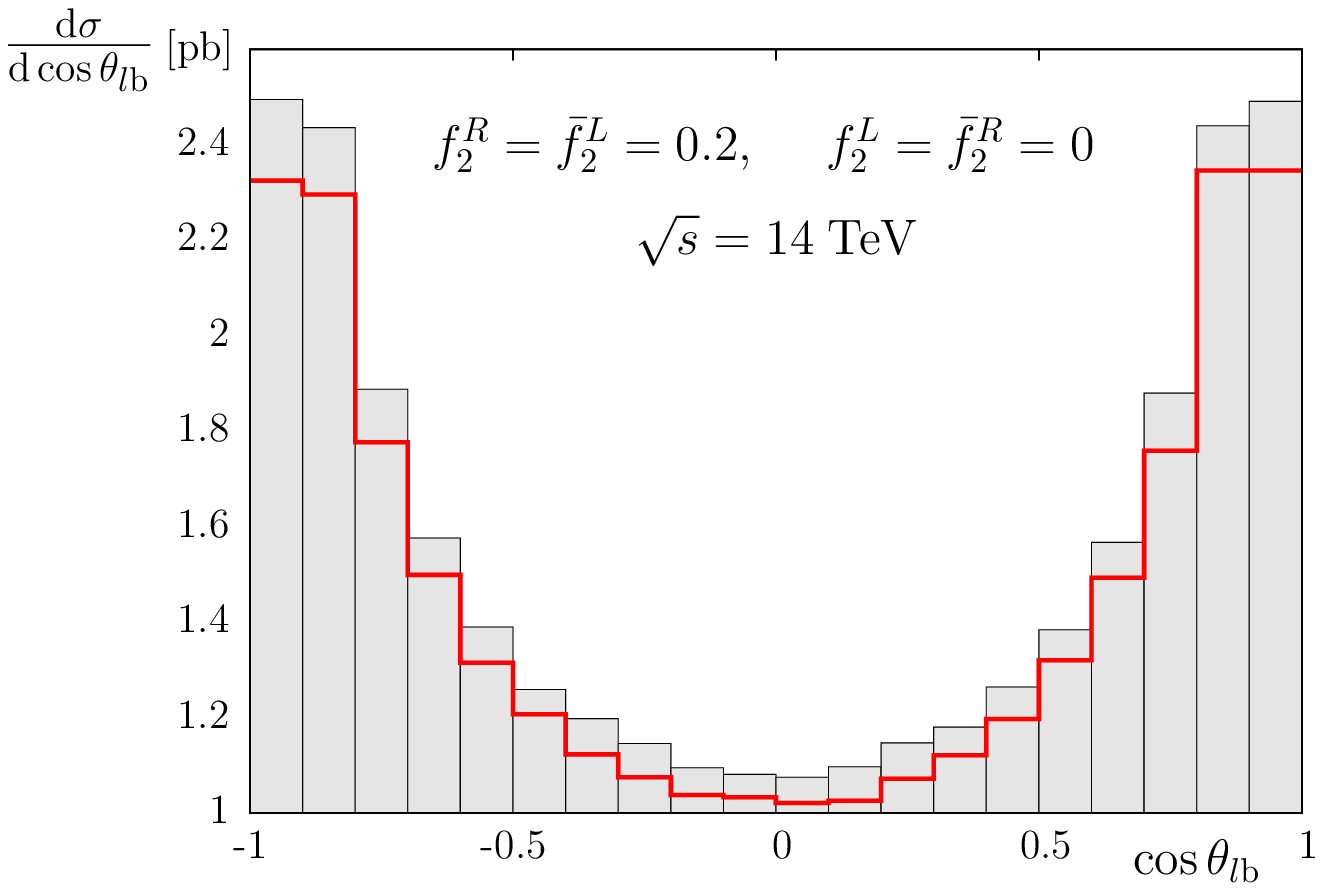}
\end{picture}\\[1.cm]
\begin{picture}(35,35)(0,0)
\includegraphics{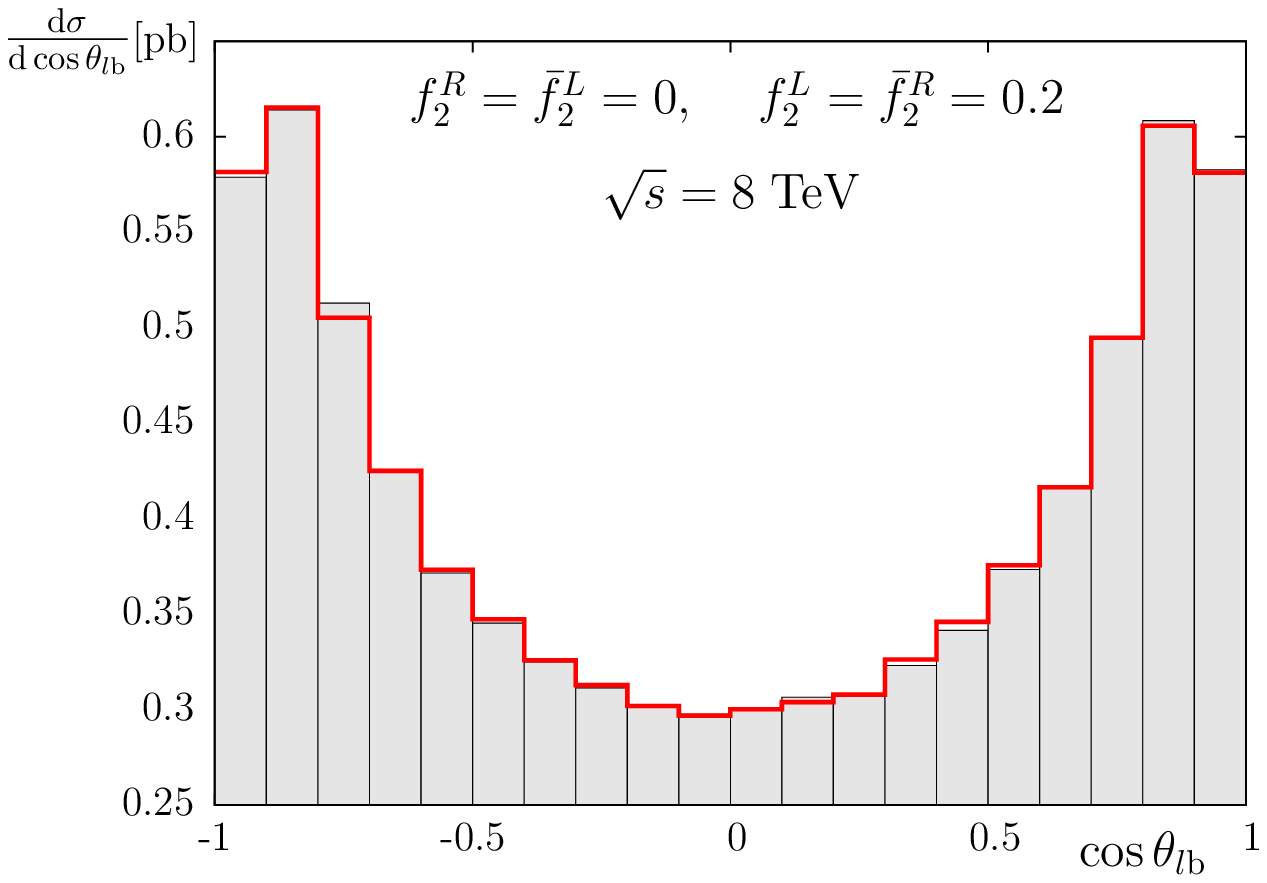}
\end{picture}
\hfill
\begin{picture}(35,35)(0,0)
\includegraphics{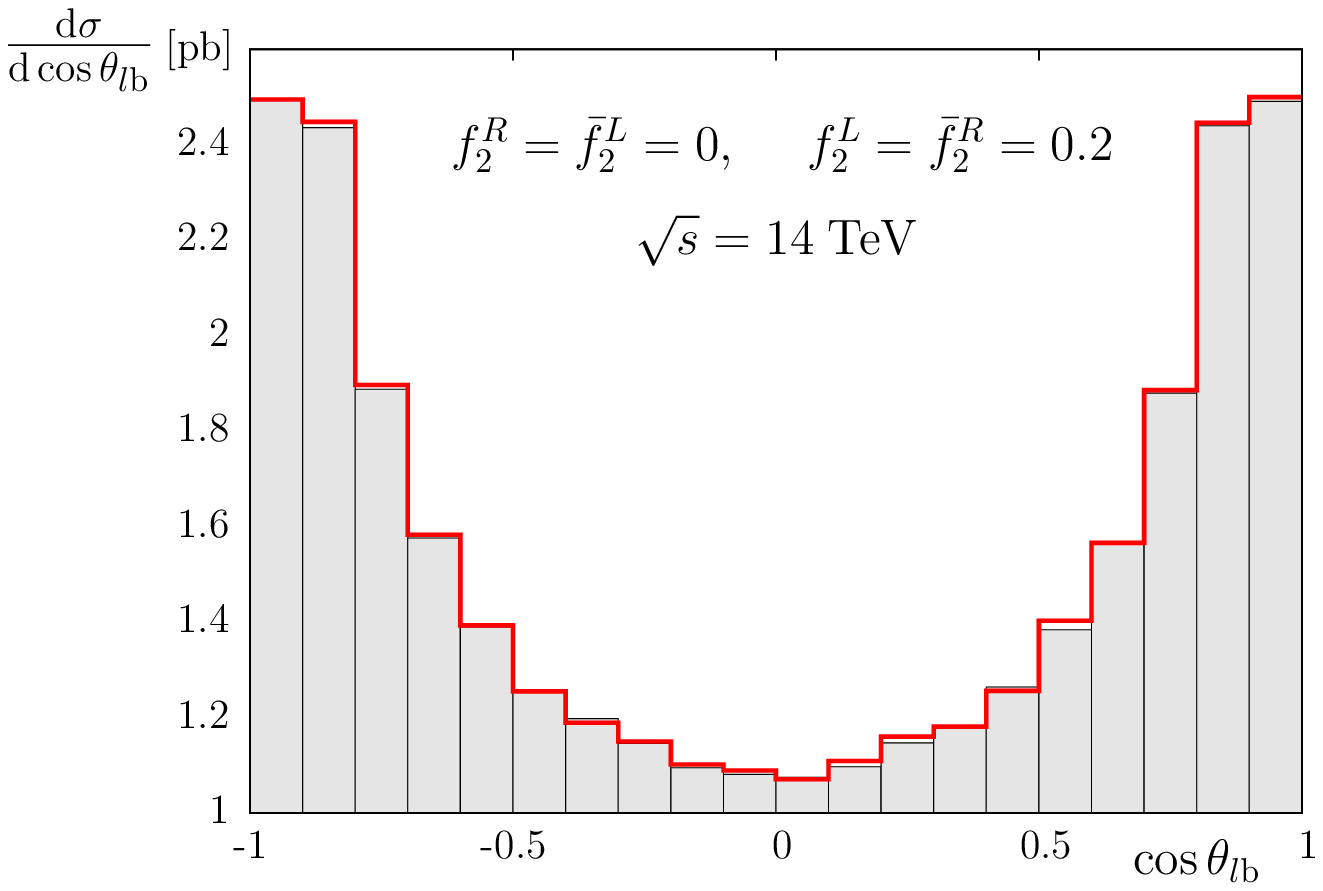}
\end{picture}\\[1.cm]
\begin{picture}(35,35)(0,0)
\includegraphics{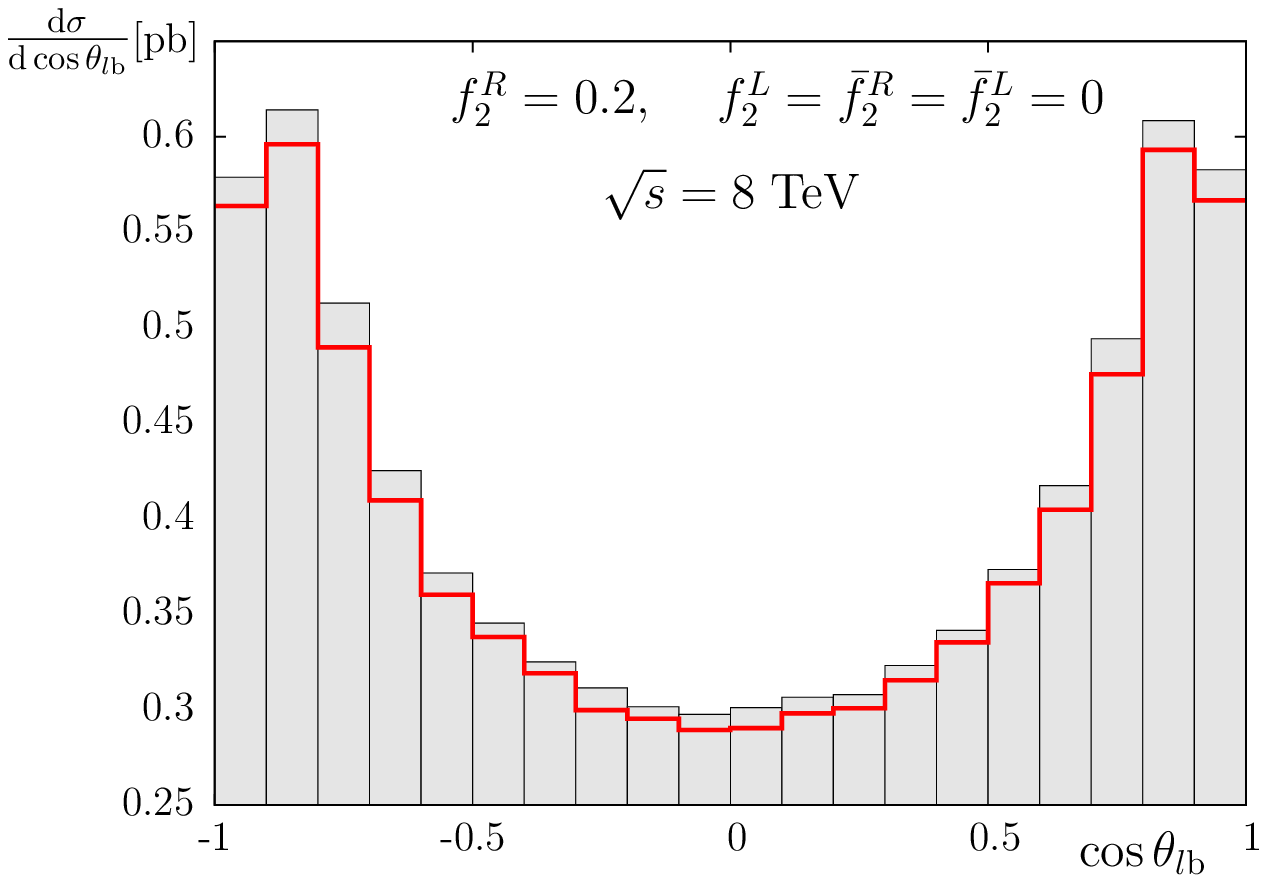}
\end{picture}
\hfill
\begin{picture}(35,35)(0,0)
\includegraphics{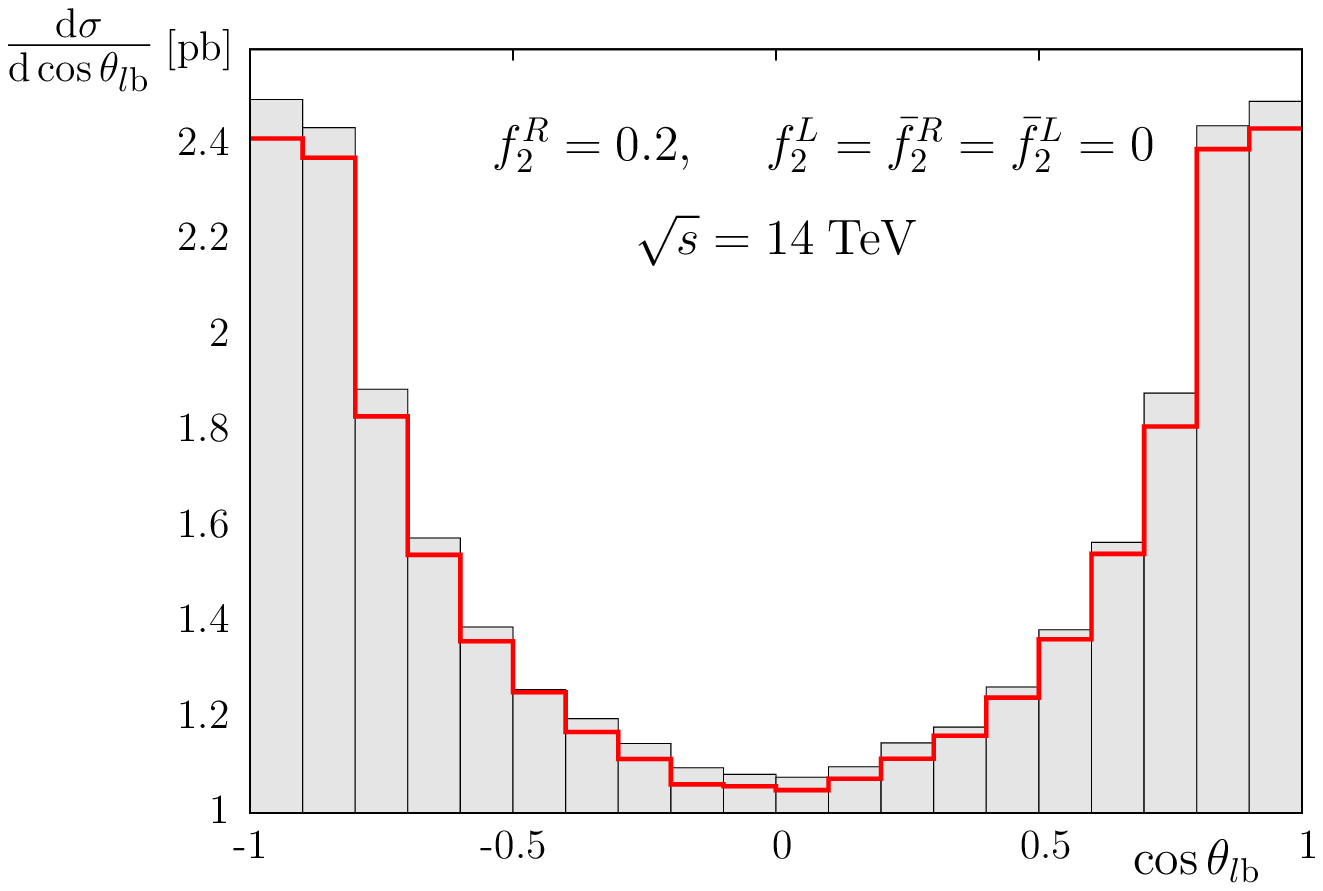}
\end{picture}\\[1.cm]
\begin{picture}(35,35)(0,0)
\includegraphics{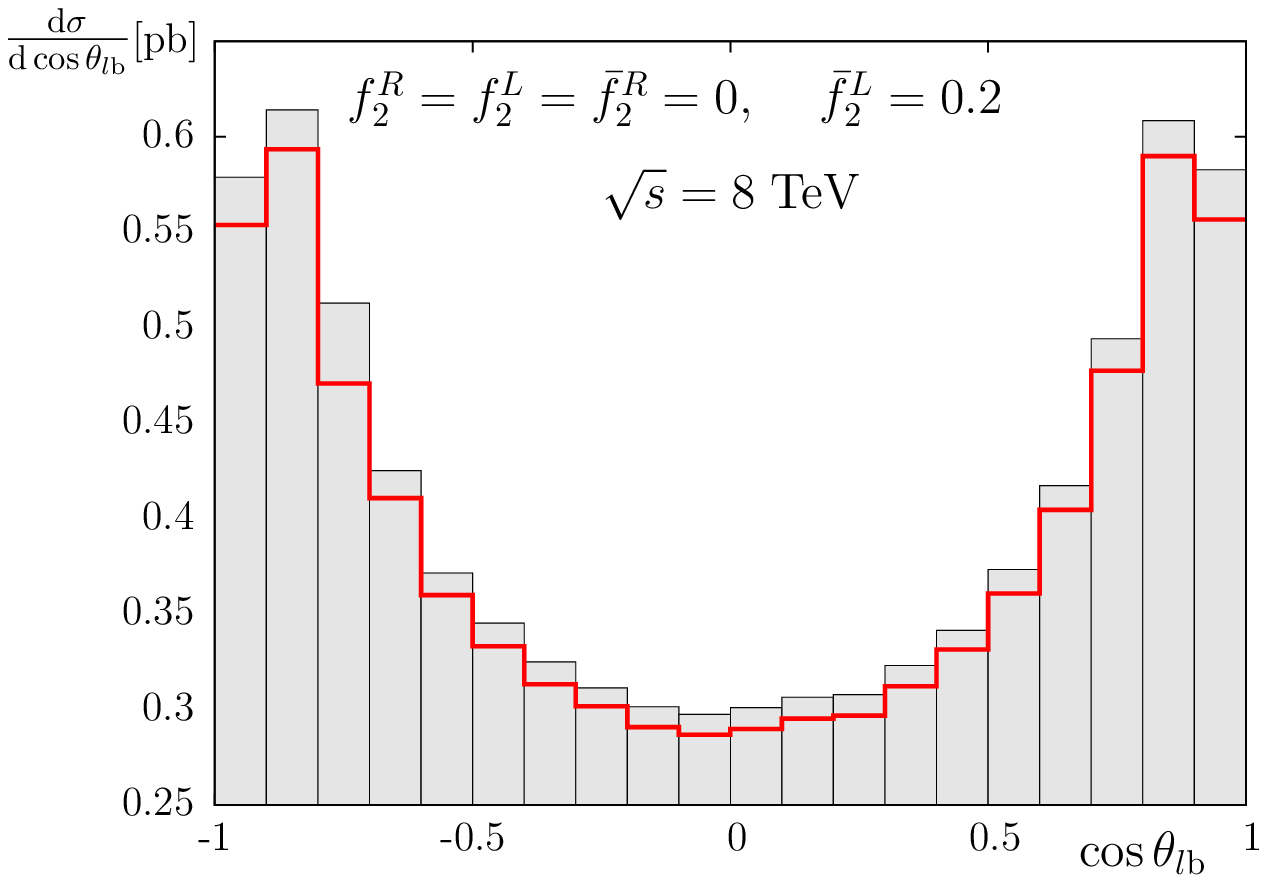}
\end{picture}
\hfill
\begin{picture}(35,35)(0,0)
\includegraphics{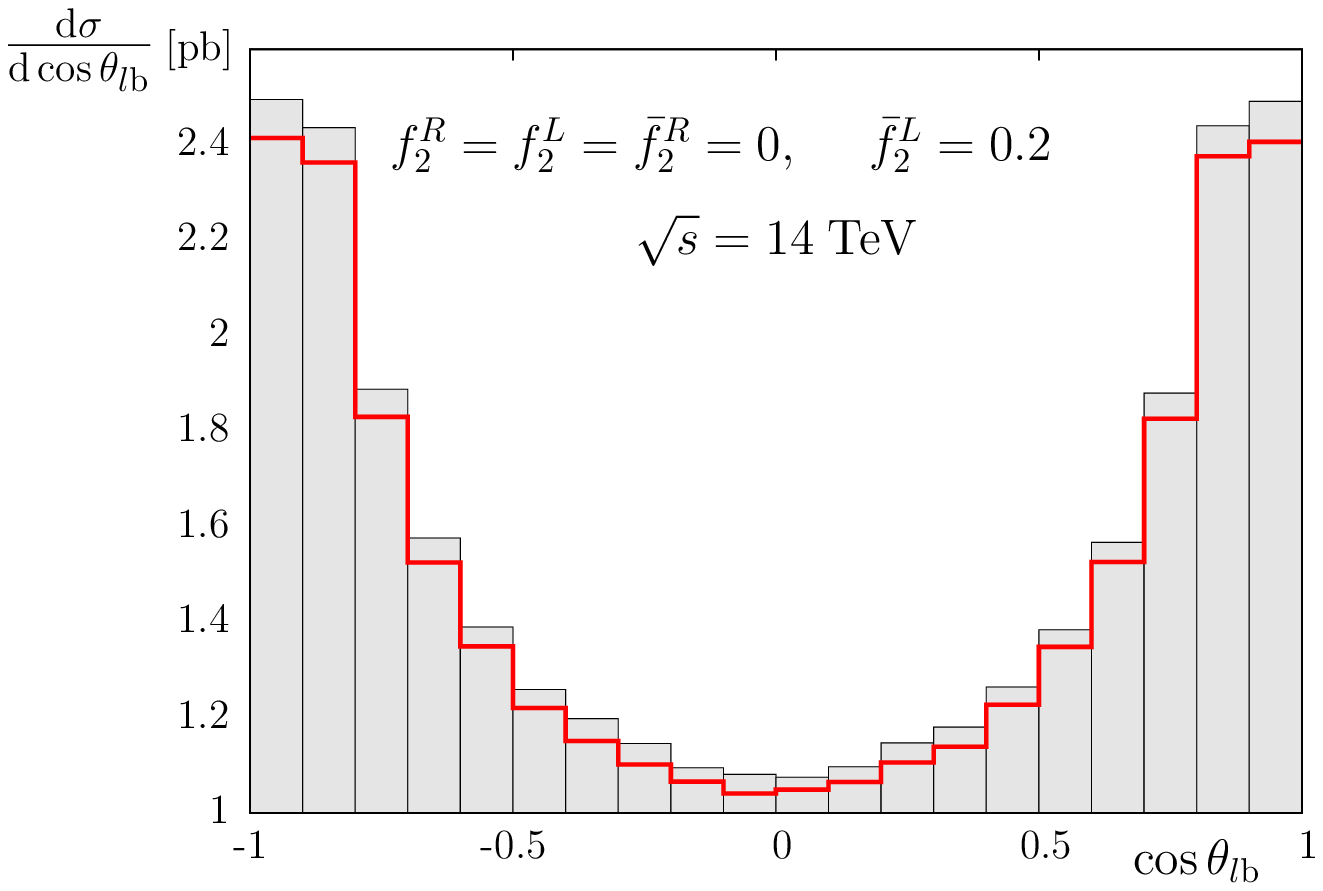}
\end{picture}
\end{center}
\vspace*{-2.cm}
\caption{Distributions in $\cos\theta_{l{\rm b}}$, with $\theta_{l{\rm b}}$ being 
an angle between the momenta of the final state $\mu^-$ of reaction 
(\ref{ppbbudmn}) and the beam,
in $pp$ collisions at $\sqrt{s}=8$~TeV (left) and $\sqrt{s}=14$~TeV 
(right) for CP-even (two upper rows) and CP-odd (two lower rows) 
choices of the tensor form factors of (\ref{lagr}).}
\label{figctlb}
\end{figure}

\begin{figure}[htb]
\vspace{100pt}
\begin{center}
\setlength{\unitlength}{1mm}
\begin{picture}(35,35)(0,0)
\includegraphics{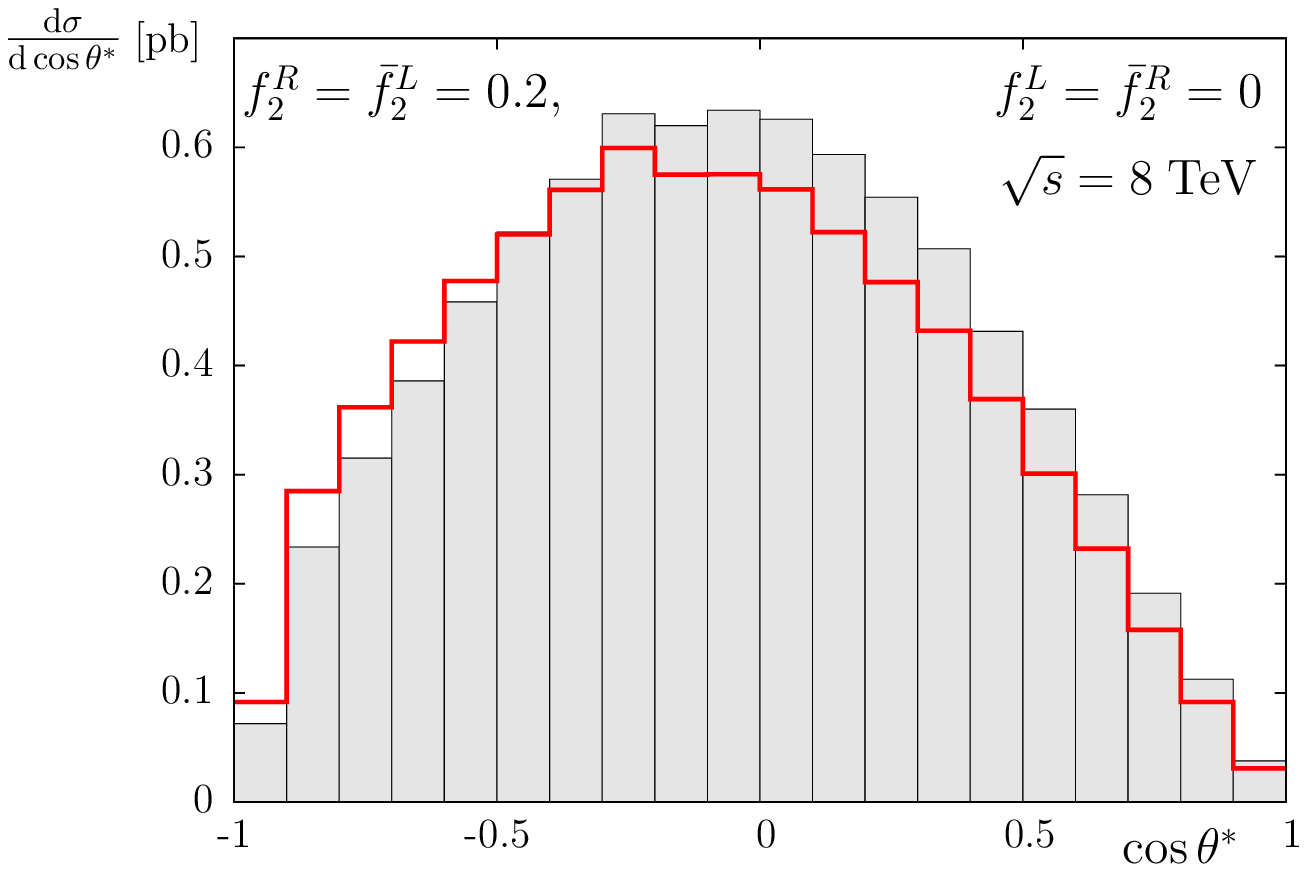}
\end{picture}
\hfill
\begin{picture}(35,35)(0,0)
\includegraphics{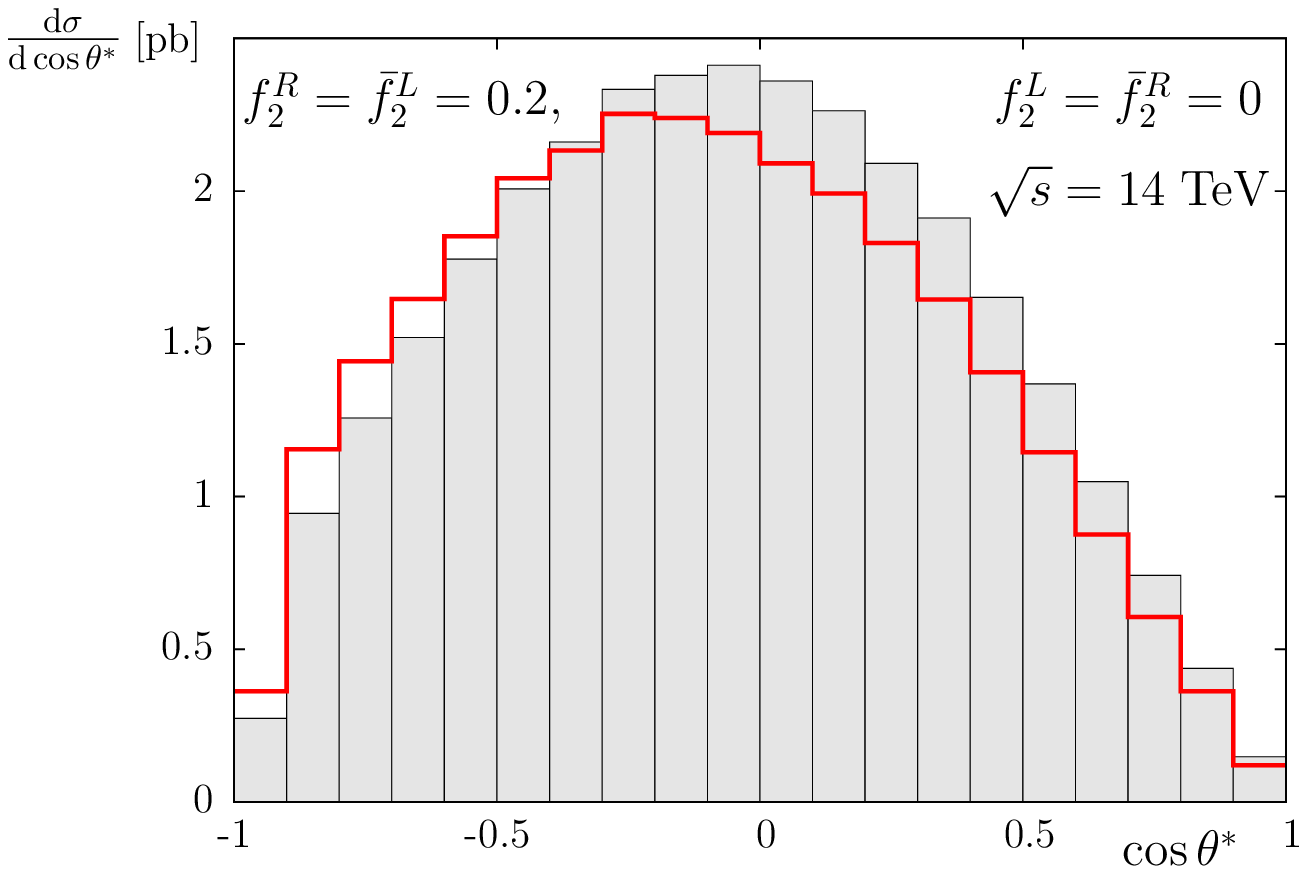}
\end{picture}\\[1.cm]
\begin{picture}(35,35)(0,0)
\includegraphics{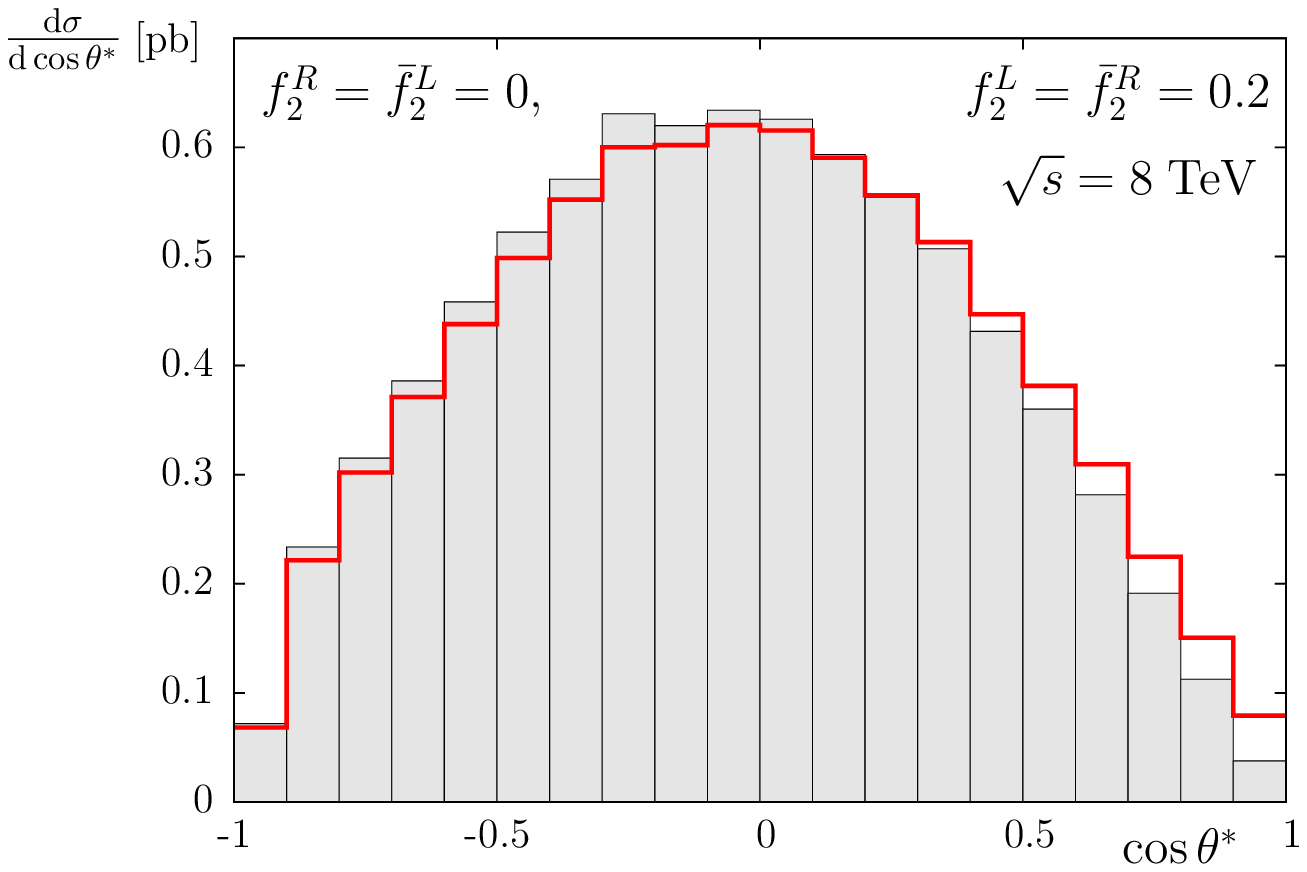}
\end{picture}
\hfill
\begin{picture}(35,35)(0,0)
\includegraphics{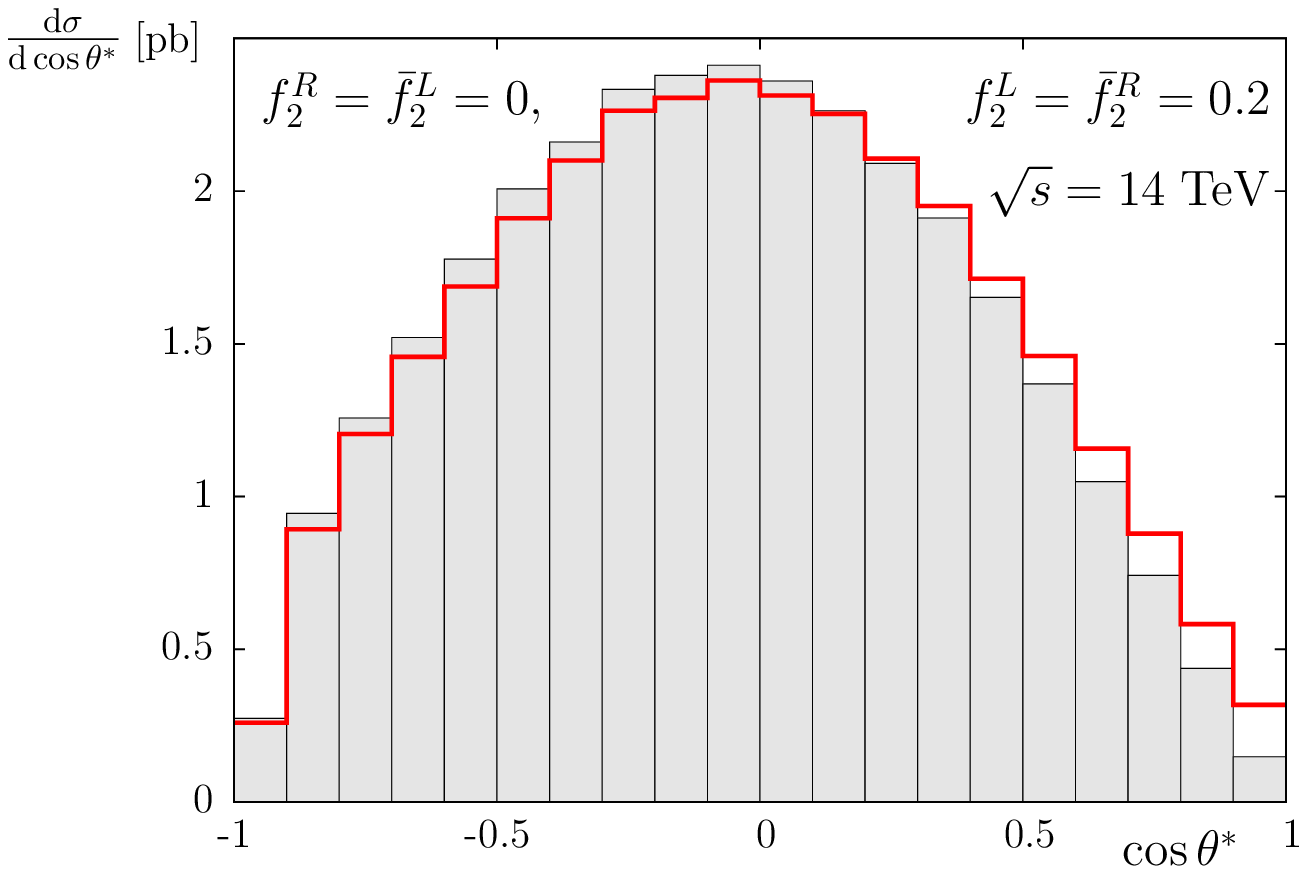}
\end{picture}\\[1.cm]
\begin{picture}(35,35)(0,0)
\includegraphics{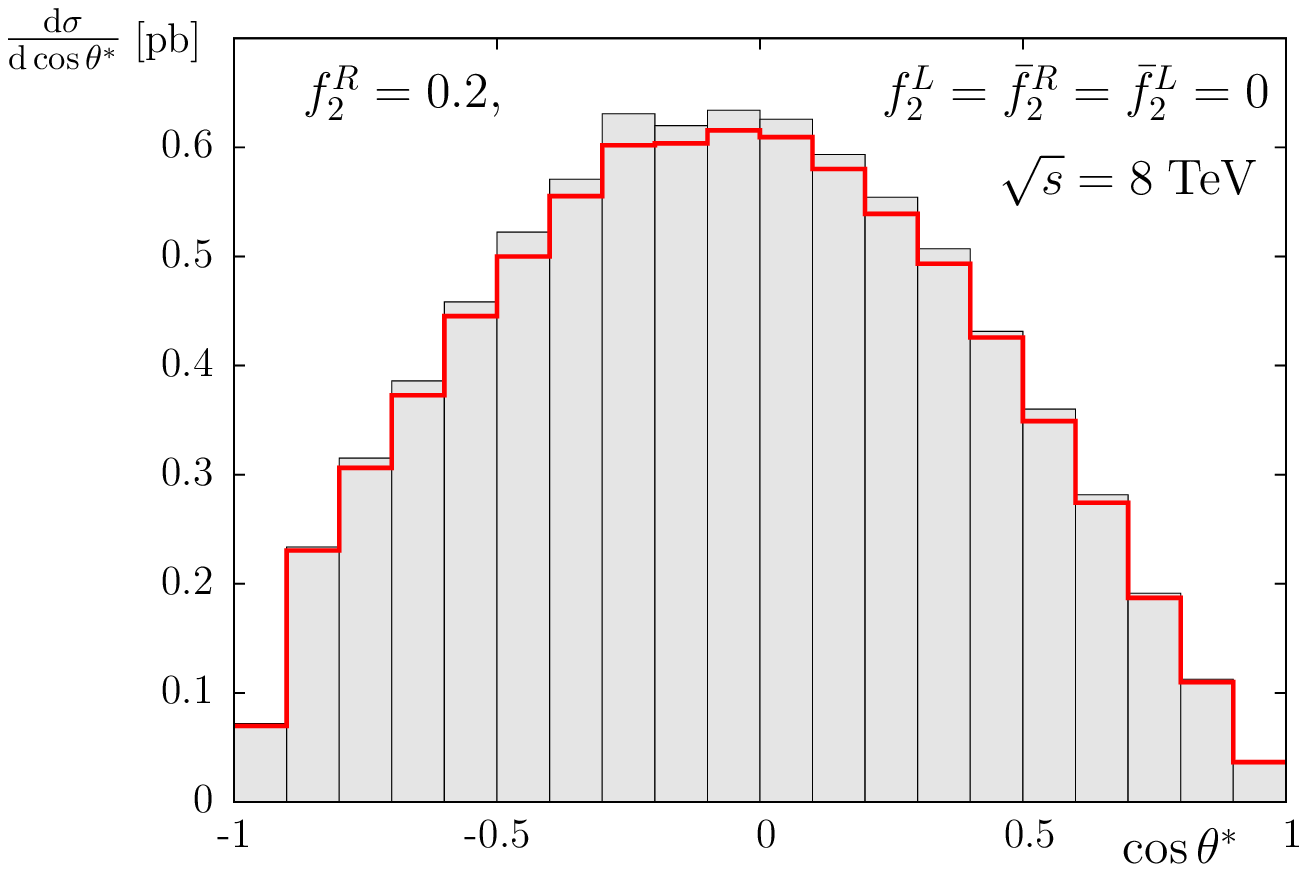}
\end{picture}
\hfill
\begin{picture}(35,35)(0,0)
\includegraphics{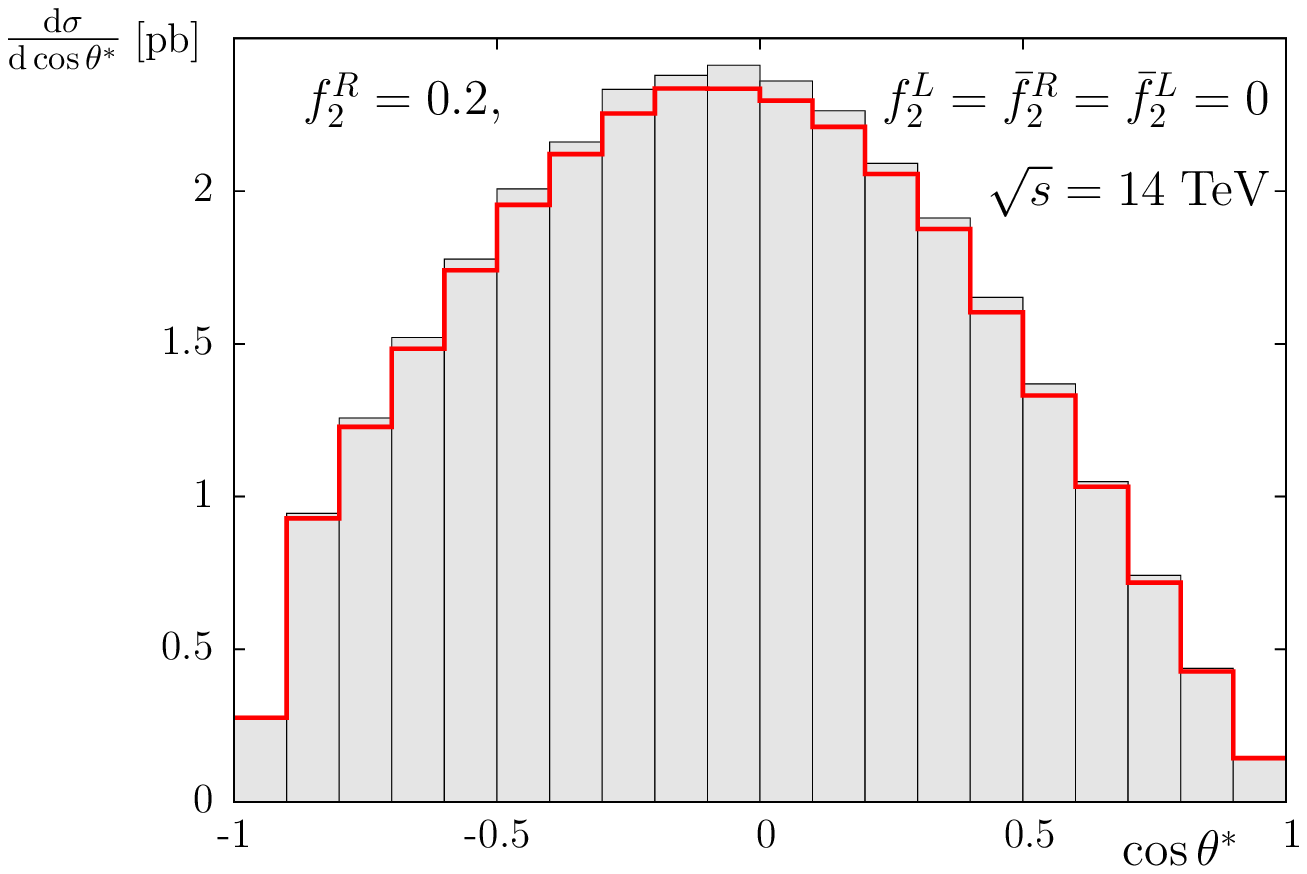}
\end{picture}\\[1.cm]
\begin{picture}(35,35)(0,0)
\includegraphics{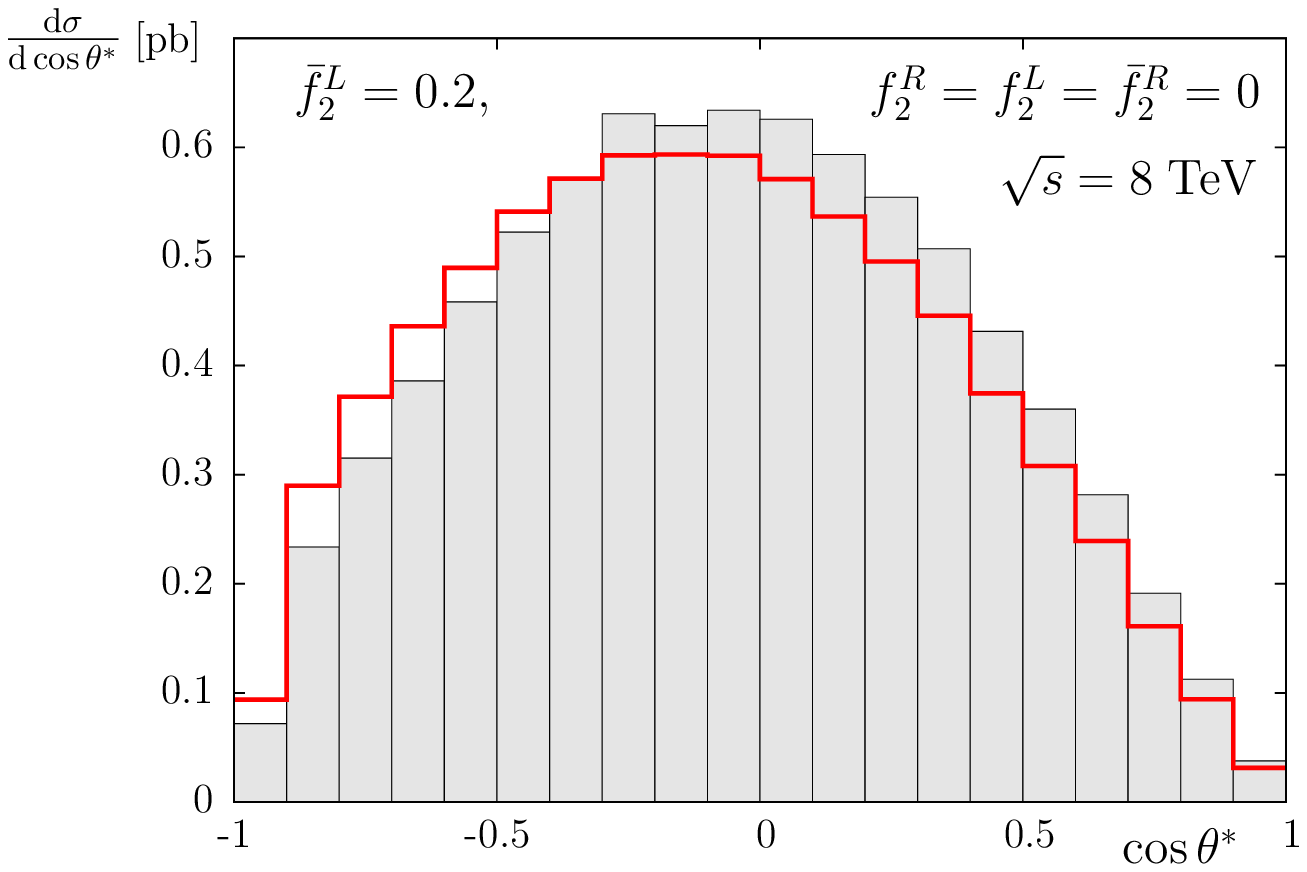}
\end{picture}
\hfill
\begin{picture}(35,35)(0,0)
\includegraphics{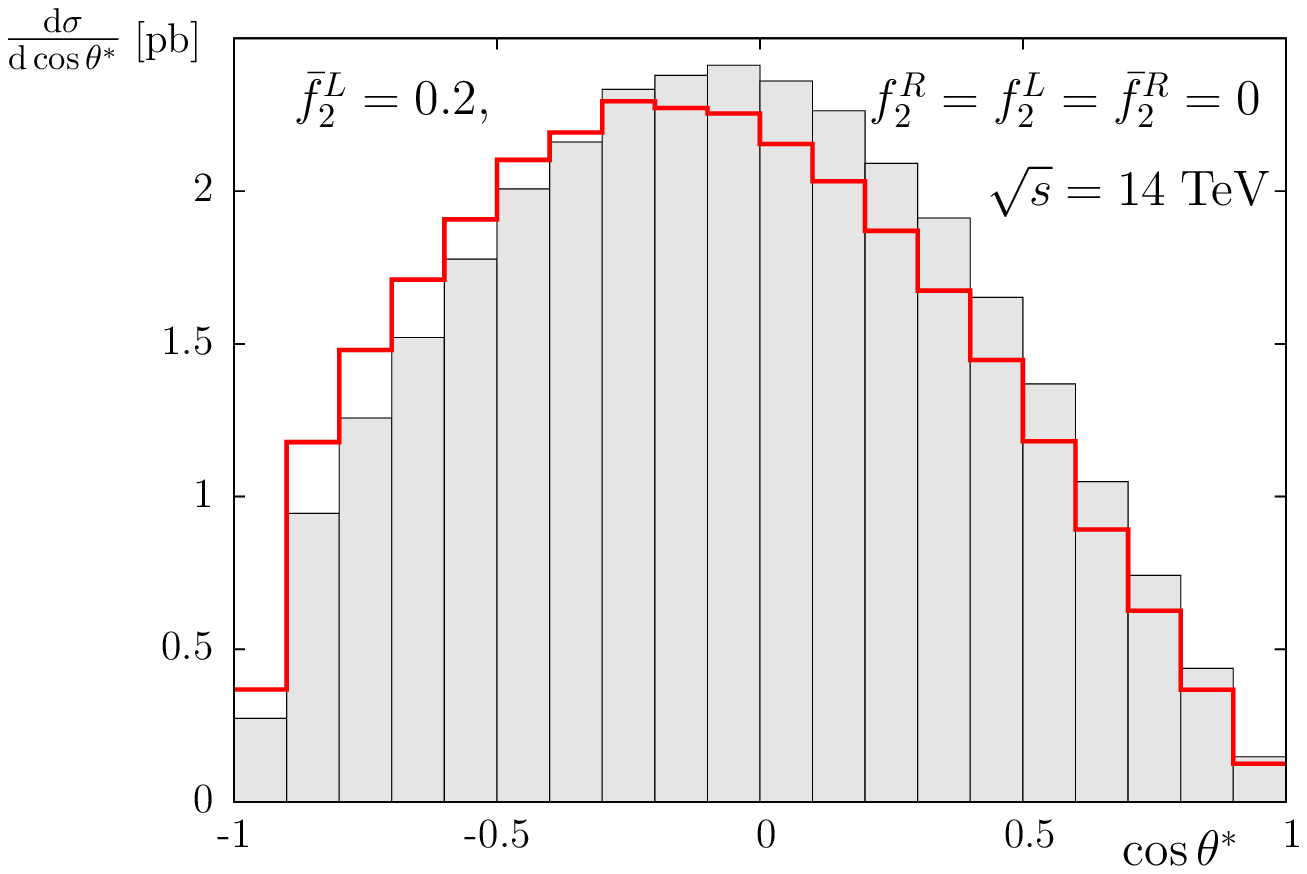}
\end{picture}
\end{center}
\vspace*{-2.cm}
\caption{Distributions in $\cos\theta^*$, where $\theta^*$ is an
angle between the momentum of $\mu^-$ and the reversed momentum of 
$\bar b$-quark of (\ref{ppbbudmn}), both boosted to the $W$-boson rest frame,  
in $pp$ collisions at $\sqrt{s}=8$~TeV  (left) and $\sqrt{s}=14$~TeV 
(right) for CP-even (two upper rows) and CP-odd (two lower rows) 
choices of the tensor form factors of (\ref{lagr}).}
\label{figcthstar}
\end{figure}

Effects of the anomalous $Wtb$ coupling must be considered relative to the
corresponding SM result. Therefore, in Figs.~\ref{figptl}--\ref{figcthstar}
the differential cross sections of (\ref{ppbbudmn}) calculated with 
different choices of the tensor form factors of Lagrangian (\ref{lagr})
and the vector form factors set to their SM values, i.e. 
$f_{1}^{L}=\bar{f}_{1}^{L}=1$ and $f_{1}^{R}=\bar{f}_{1}^{R}=0$, are each time 
plotted together with the corresponding SM cross section.
The actual values of the tensor form factors used are always specified in a 
plot.
The absolute size of the cross sections should be treated with great care,
as it to large degree depends on the choice of PDFs
or factorization scale. To diminish the dependence on the latter higher order
QCD corrections should be taken into account. It should be also realized that
the form factors get contributions from the EW loop corrections
that, however, should be much smaller than the value of 0.2 used in the plots.
The relative size of the modification is on the other hand independent of
the choice of PDFs which has been checked explicitly by repeating the 
calculations with MSTW PDFs \cite{MSTW}.

Figs.~\ref{figptl}--\ref{figcthstar} show the results for a few distributions 
of the final state $\mu^-$ 
of reaction (\ref{ppbbudmn}) at two different $pp$ collision energies
of 8~TeV and 14~TeV. In each of the figures, the SM cross
section is plotted with grey boxes and the cross sections
in the presence of different CP-even (two upper rows) 
and CP-odd (two lower rows) choices of the tensor form factors of 
Lagrangian (\ref{lagr}) are plotted with solid lines.
The transverse momentum and rapidity distributions of the 
$\mu^-$ are shown in Figs.~\ref{figptl} and \ref{figrapl}.
The distributions in $\cos\theta_{l{\rm b}}$, where $\theta_{l{\rm b}}$ 
is the angle between the momentum of $\mu^-$ in the $pp$ centre of mass frame
and the beam, are plotted in Fig.~\ref{figctlb}.
Finally, the distributions in $\cos\theta^*$, where $\theta^*$ is
the angle between the momentum of $\mu^-$ and the reversed momentum 
of the $b$-quark, both boosted to the rest frame of the $W$-boson,
are plotted in Fig.~\ref{figcthstar}.
Distributions in $\cos\theta^*$ are usually used in order to determine
the helicity fractions of the $W$-boson produced in top quark decays, see, 
e.g., \cite{wtbATLAS}, \cite{wtbCMS}.

Changes in the distributions are rather moderate, at most of the order of 
several per cent, both for the CP-even and CP-odd combinations 
of the tensor form factors.
This observation can be regarded as another example of 
the decoupling theorem of \cite{GHR} that was originally 
proven in the NWA, but now seems to work also for
the off shell top quark pair production and decay in proton-proton collisions
at the LHC energies.
Although all the cross sections increase by about a factor 4 
if the energy of $pp$ collisions is increased from
$\sqrt{s}=8$~TeV to  $\sqrt{s}=14$~TeV, the relative changes caused
by the tensor form factors are fairly independent of the $pp$ collision energy,
as can be seen by comparing plots on the left- and right-hand sides of
Figs.~\ref{figptl}--\ref{figcthstar}.
Let us note  that shapes of the distributions remain practically 
unchanged by the non zero form factors, both in the CP-even
and CP-odd case, except for the distribution in $\cos\theta^*$ of 
Fig.~\ref{figcthstar}.

\begin{figure}[htb]
\vspace{100pt}
\begin{center}
\setlength{\unitlength}{1mm}
\begin{picture}(35,35)(0,0)
\includegraphics{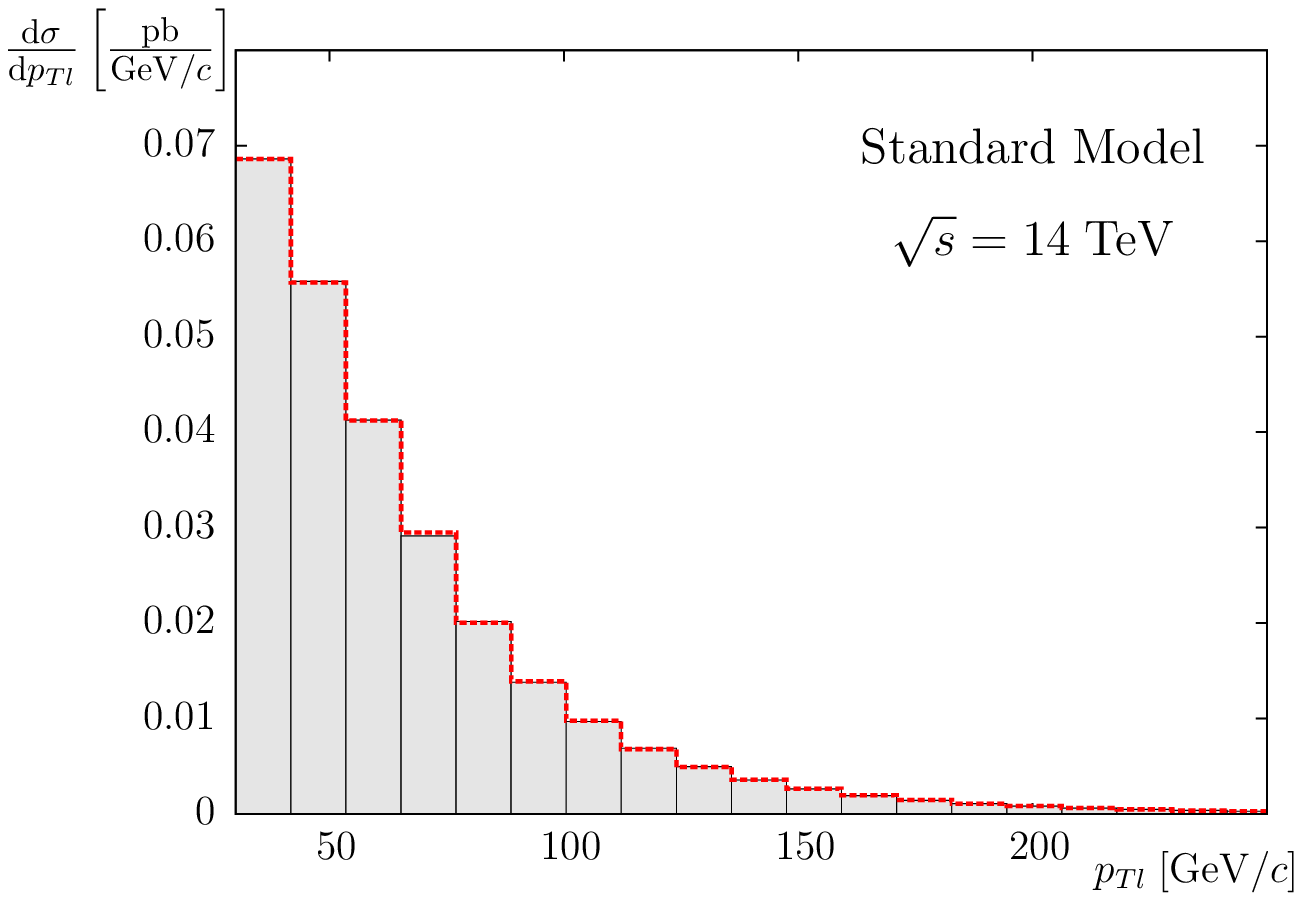}
\end{picture}
\hfill
\begin{picture}(35,35)(0,0)
\includegraphics{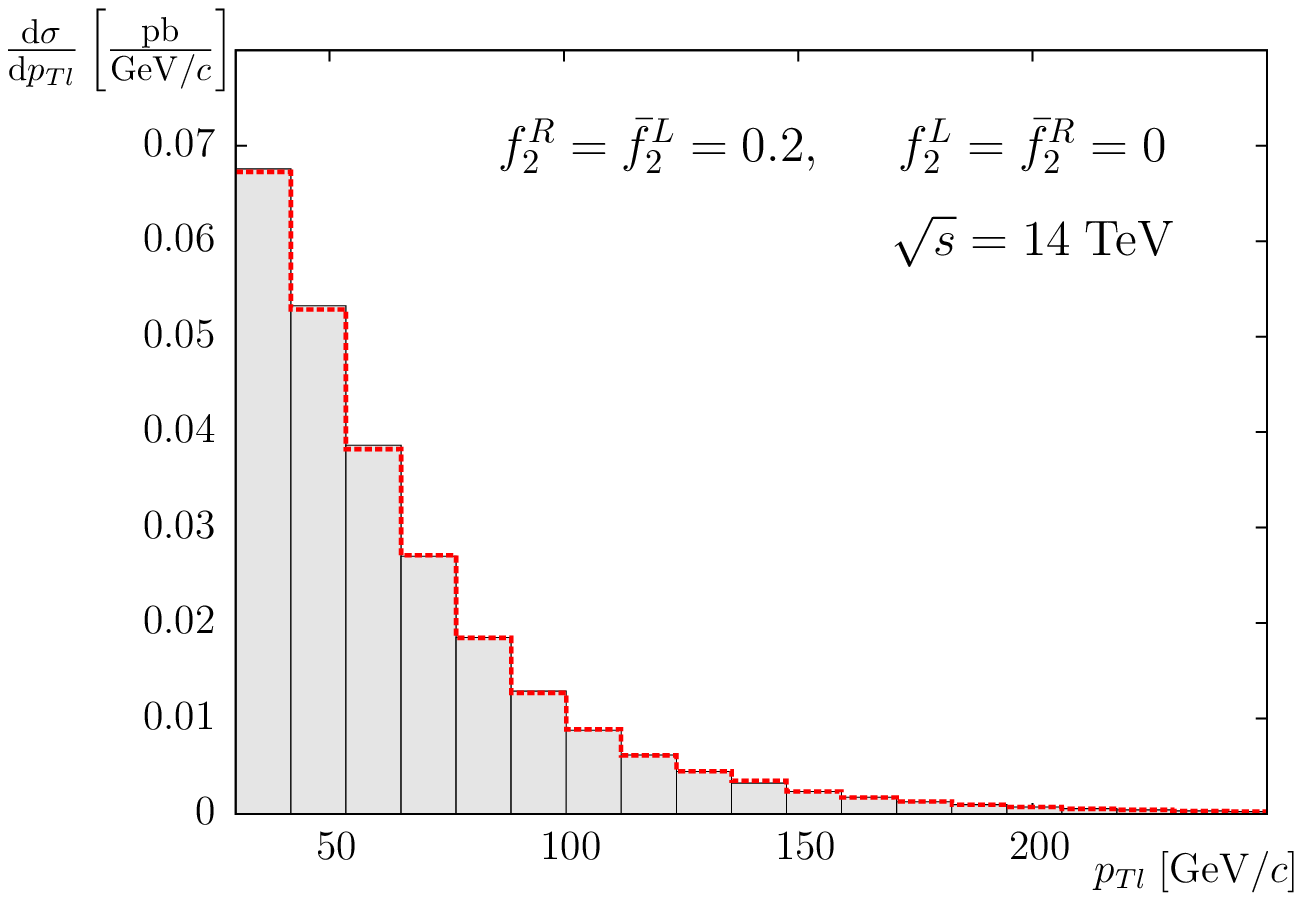}
\end{picture}\\[1.cm]
\begin{picture}(35,35)(0,0)
\includegraphics{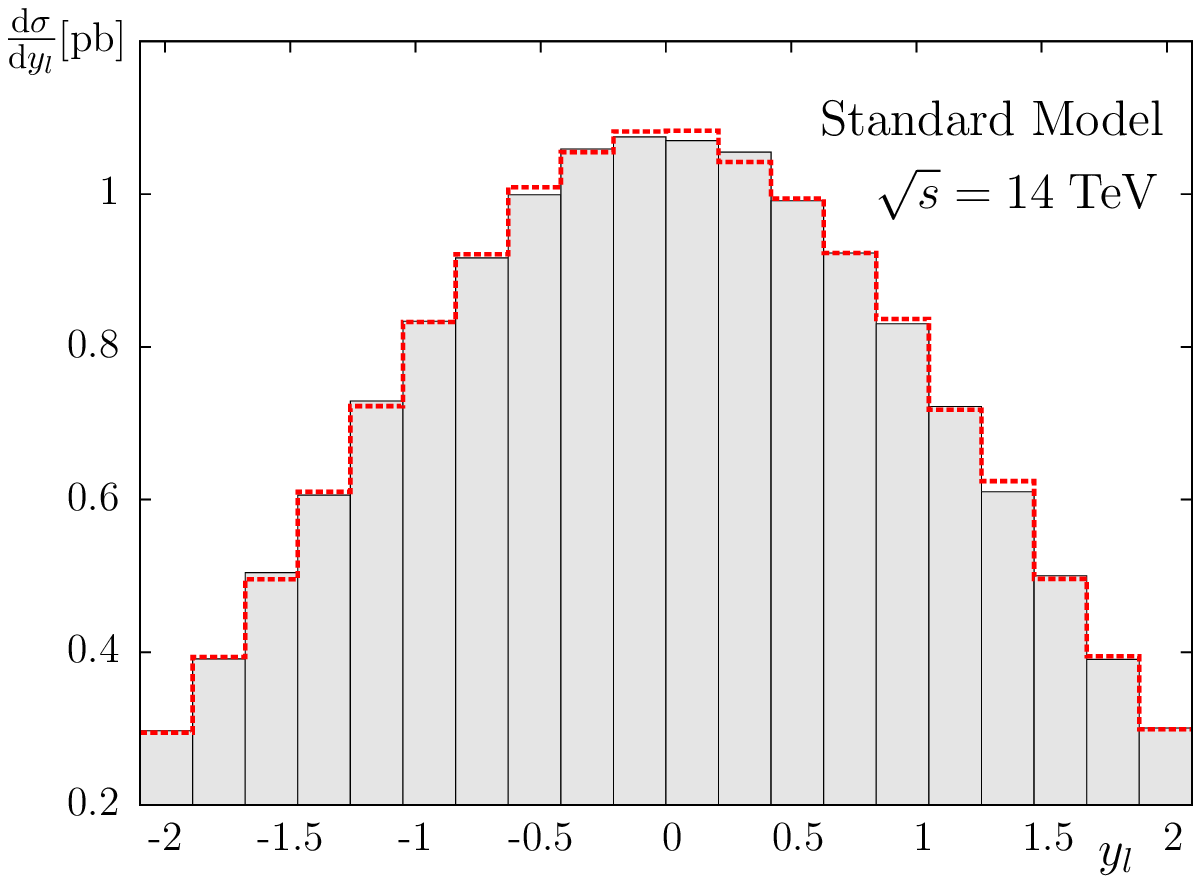}
\end{picture}
\hfill
\begin{picture}(35,35)(0,0)
\includegraphics{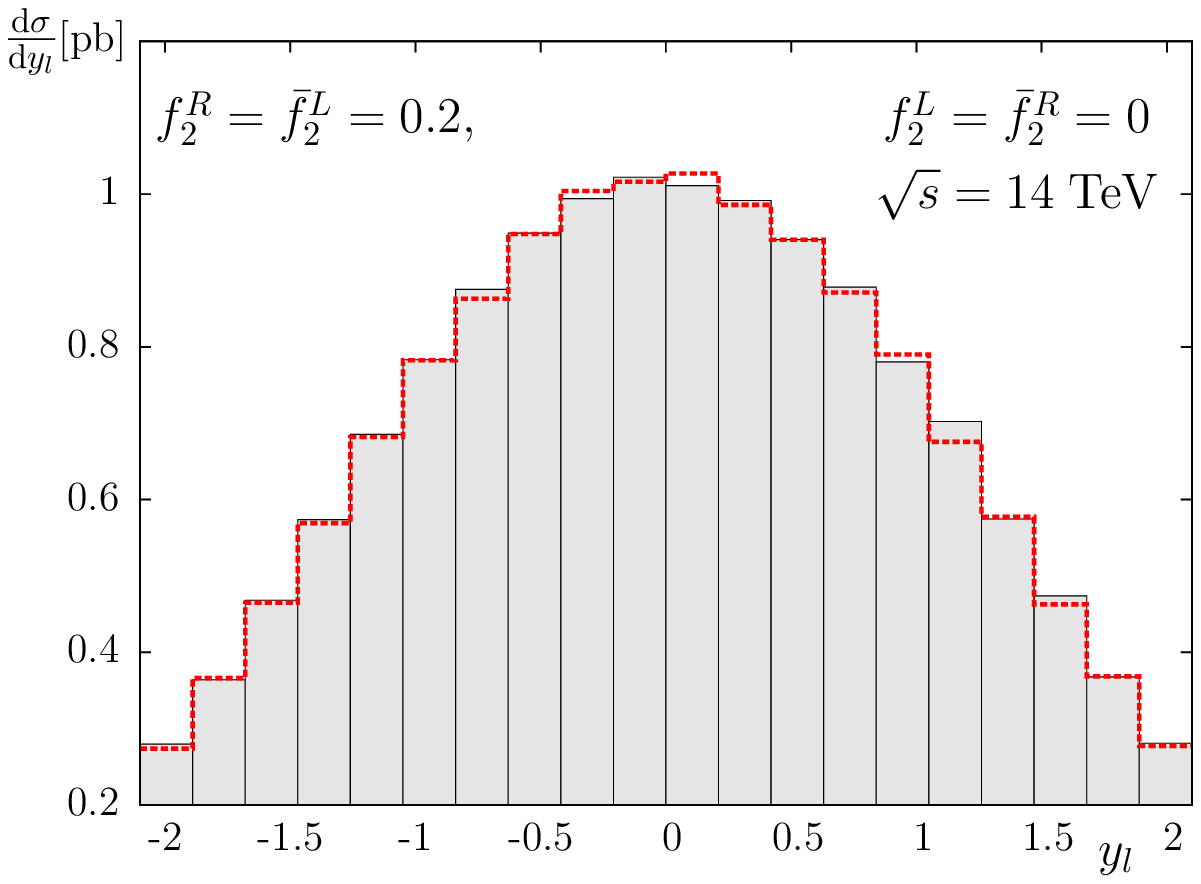}
\end{picture}\\[1.cm]
\begin{picture}(35,35)(0,0)
\includegraphics{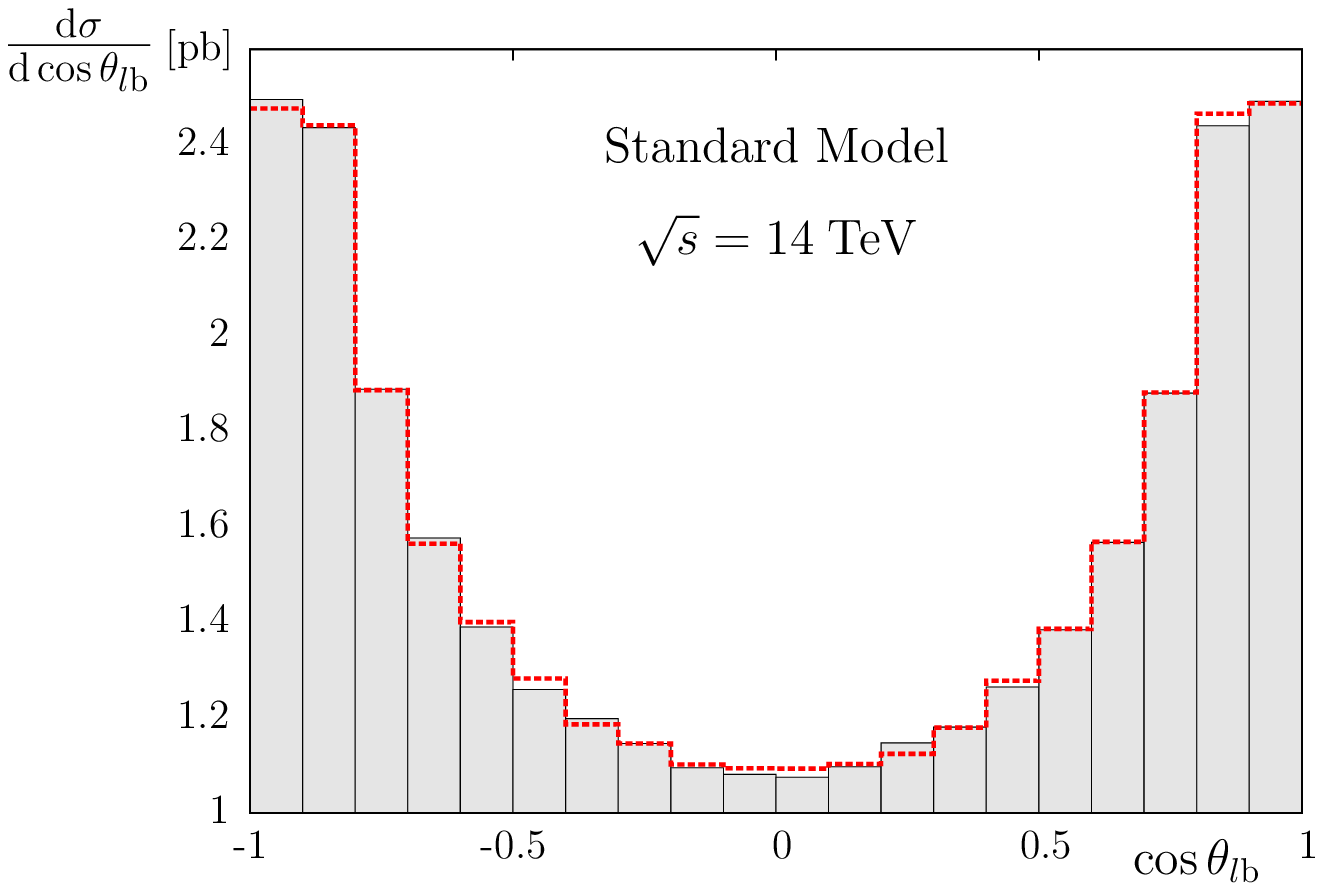}
\end{picture}
\hfill
\begin{picture}(35,35)(0,0)
\includegraphics{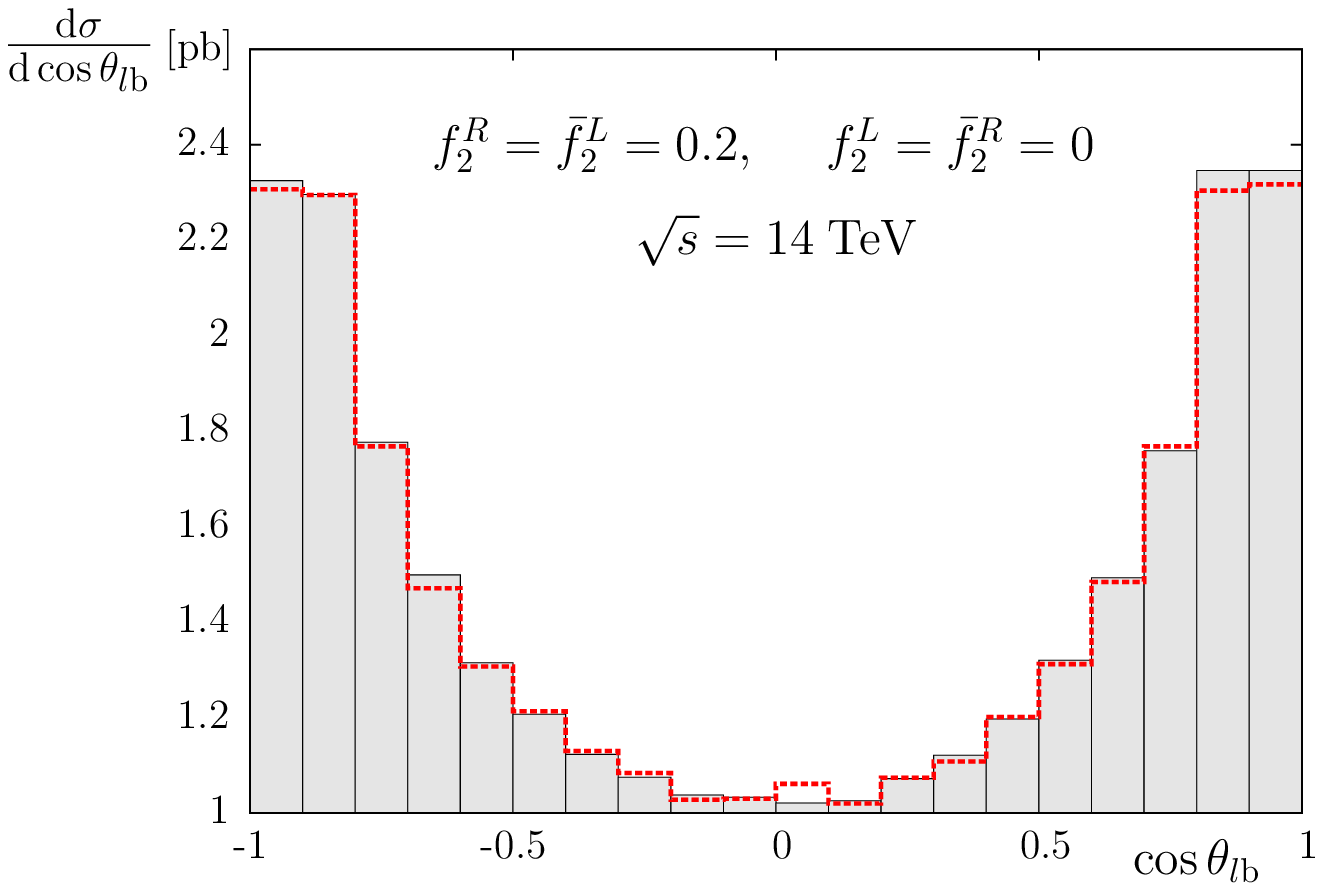}
\end{picture}\\[1.cm]
\begin{picture}(35,35)(0,0)
\includegraphics{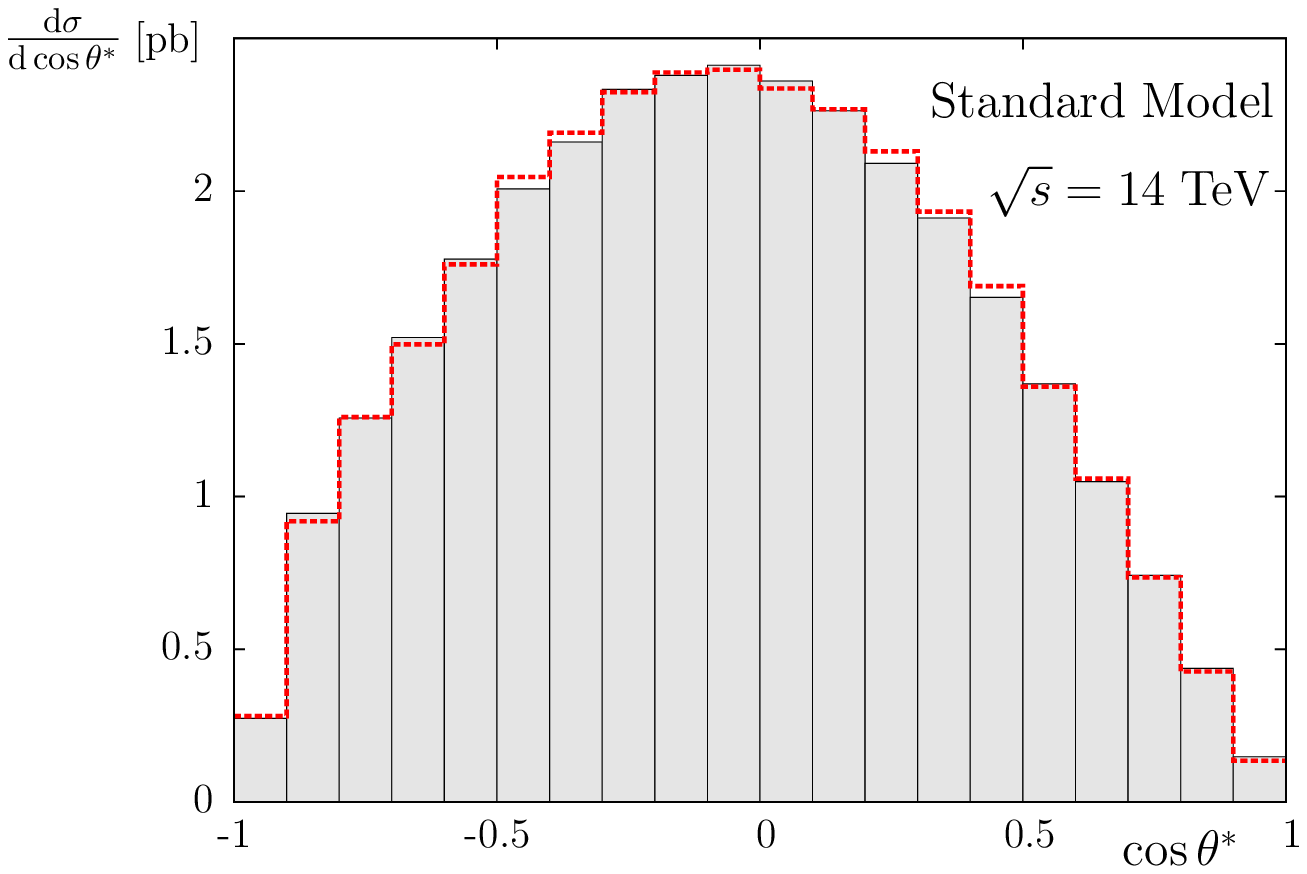}
\end{picture}
\hfill
\begin{picture}(35,35)(0,0)
\includegraphics{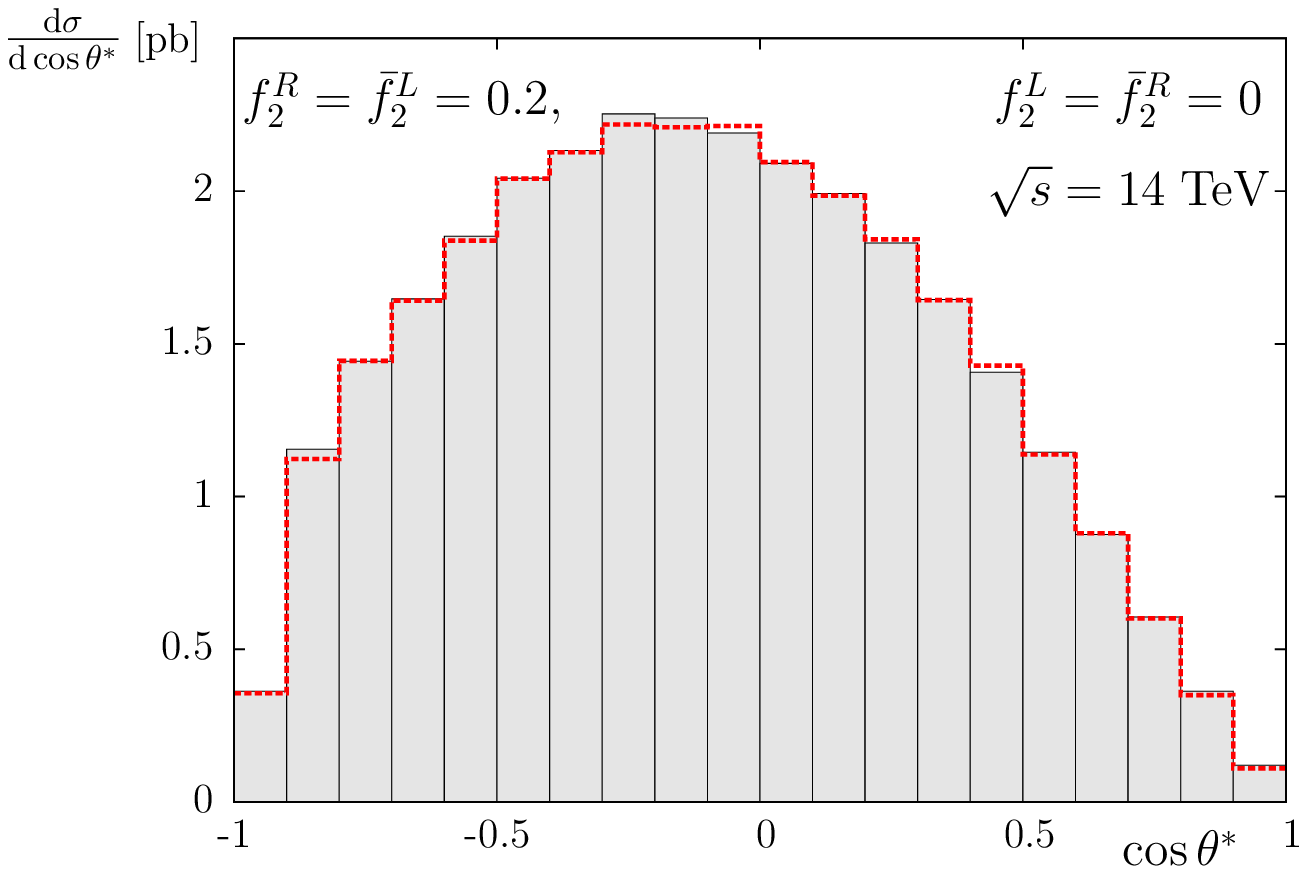}
\end{picture}
\end{center}
\vspace*{-2.cm}
\caption{Distributions  of $\mu^-$ of (\ref{ppbbudmn}) at $\sqrt{s}=14$~TeV
in the same variables as in Figs.~\ref{figptl}--\ref{figcthstar}.
The distributions computed with the full set of leading order
Feynman diagrams are plotted with the shaded boxes and those computed with
the $t\bar t$ production signal diagrams only are plotted with the dashed 
lines.}
\label{figsva}
\end{figure}

The distributions  of $\mu^-$ of reaction (\ref{ppbbudmn}) at $\sqrt{s}=14$~TeV
in the same variables as in Figs.~\ref{figptl}--\ref{figcthstar}
computed with the full set of leading order
Feynman diagrams, plotted with the shaded boxes, and with
the $t\bar t$ production signal diagrams, plotted with the dashed 
lines, are compared in Fig.~\ref{figsva}. 
The comparison demonstrates very
little effect of the off resonance background contributions
in the presence of acceptance cuts (\ref{cuts}).
The results for other combinations of the tensor form factors, which are
not shown, look very similarly.


\section{Summary}

A new version of {\tt carlomat} \cite{carlomat2}, a general purpose program 
for the MC computation of lowest order cross sections,
has been used to compute the transverse momentum, rapidity and 
two angular distributions of the final state $\mu^-$ of reaction 
(\ref{ppbbudmn}) in the presence 
of the anomalous $Wtb$ coupling with operators of dimension up to 
five. The considered CP-even and CP-odd combinations of the tensor
form factors have
rather small effect on the distributions which actually could be expected, as
the top quarks are produced unpolarized. 
At the same time, the shapes of the presented distributions remain practically
unchanged, except for the distribution
in $\cos\theta^*$, where $\theta^*$ is the angle between the momentum 
of $\mu^-$ and the reversed momentum of the $b$-quark, both
boosted to the rest frame of $W$-boson. It is just the change in shape
of the $\cos\theta^*$ distribution that potentially gives the best prospects 
for improving limits on the tensor form factors in future measurements 
at the LHC.

It has been also shown that the off resonance background contributions
have rather little impact on the distributions independently of
whether the anomalous tensor form factors are present or not.

{\bf Acknowledgements:} 
This project was supported in part with financial resources 
of the Polish National Science Centre (NCN) under grant decision 
number DEC-2011/03/B/ST6/01615 and by the Research Executive 
Agency (REA) of the European Union under the Grant Agreement number 
PITN-GA-2010-264564 (LHCPhenoNet).

\end{document}